\documentclass[sigconf]{acmart}
\usepackage{booktabs} 
\usepackage{mdframed}
\usepackage{framed}
\usepackage{hyphenat}
\usepackage{algpseudocode}
\usepackage{amsmath,epsfig}
\usepackage[boxed,ruled,vlined,linesnumbered]{algorithm2e}
\usepackage{amsthm}
\usepackage{setspace}
\usepackage{enumitem}
\usepackage{ragged2e}
\usepackage{multirow}
\usepackage{hhline}
\usepackage{wrapfig}
\usepackage{amssymb}
\usepackage{caption}
\usepackage{subcaption}
\usepackage{graphicx}
\usepackage{amsmath,epsfig}
\usepackage{amsthm}

\newtheorem{attackgame}{ATTACK GAME}

\newcommand{\B}{\vspace*{-\smallskipamount}}
\newcommand{\BB}{\vspace*{-\medskipamount}}
\newcommand{\BBB}{\vspace*{-\bigskipamount}}
\newcommand{\xor}{\mathbin{\oplus}}

\newcommand{\aaa}{\mathcal{A}}

\newcommand{\aexp}{\mathrm{\mathbf{Exp}}_{\aaa}}

\copyrightyear{2019}
\acmYear{2019}
\setcopyright{acmcopyright}
\acmConference[CODASPY '19]{Ninth ACM Conference on Data and Application Security and Privacy}{March 25--27, 2019}{Richardson, TX, USA}
\acmBooktitle{Ninth ACM Conference on Data and Application Security and Privacy (CODASPY '19), March 25--27, 2019, Richardson, TX, USA}
\acmPrice{15.00}
\acmDOI{10.1145/3292006.3300043}
\acmISBN{978-1-4503-6099-9/19/03}

\begin{document}

\title{Verifiable Round-Robin Scheme for Smart Homes}\titlenote{\textbf{Accepted in ACM Conference on Data and Application Security and Privacy (CODASPY), 2019.} \newline This material is based on research sponsored by DARPA under agreement number FA8750-16-2-0021. The U.S. Government is authorized to reproduce and distribute reprints for Governmental purposes notwithstanding any copyright notation thereon. The views and conclusions contained herein are those of the authors and should not be interpreted as necessarily representing the official policies or endorsements, either expressed or implied, of DARPA or the U.S. Government. This work is partially supported by NSF grants 1527536 and 1545071.}

\author{Nisha Panwar, Shantanu Sharma, Guoxi Wang, Sharad Mehrotra, and Nalini Venkatasubramanian}
\affiliation{\institution{University of California, Irvine, USA.}}

 \renewcommand{\shortauthors}{Panwar et al.}
\renewcommand{\shorttitle}{Verifiable Round-Robin Scheme for Smart Homes}

\begin{abstract}
Advances in sensing, networking, and actuation technologies have resulted in the IoT wave that is expected to revolutionize all aspects of modern society. This paper focuses on the new challenges of privacy that arise in IoT in the context of smart homes. Specifically, the paper focuses on preventing the user's privacy via inferences through channel and in-home device activities. We propose a method for securely scheduling the devices while decoupling the device and channels activities. The proposed solution avoids any attacks that may reveal the coordinated schedule of the devices, and hence, also, assures that inferences that may compromise individual's privacy are not leaked due to device and channel level activities. Our experiments also validate the proposed approach, and consequently, an adversary cannot infer device and channel activities by just observing the network traffic.
 \end{abstract}


 \maketitle


\section{Introduction}
\label{sec:introduction}
The IoT devices are quickly becoming a pervasive and integral part of modern smart homes~\cite{bren}. The homeowner, typically, possesses a heterogeneous set of devices ranging from wearable devices, information/entertainment devices to smart home appliances. These devices provide comfort/assisted-living and/or improve sustainability, reduce costs, and reduce carbon footprint. For example, a Belkin Wemo switch can automatically switch lights on/off and open/close window shades based on the sunlight and time of the day. Likewise, dampers in the AC vent can be partially/fully opened/closed to modulate airflow. Other devices popular in smart homes include Nest cameras, smart door locks, Lenovo Smart Assistant, Amazon Echo, Echo dot, Echo show, Alexa, Philips-Hue Bloom/Lightstrip Plus, SteriGrip self-cleaning door handles, Unico smartbrush, Sensus Metering Systems, and Logitech Circle 2 among others.

While the emerging smart home devices provide significant benefits, the support for security in such devices is often limited to the security offered by the original equipment manufacturer (OEM). Lack of strong end-to-end architecture for security has led to devices being vulnerable to a variety of attacks. For example, McAfee Labs~\cite{guoxi1} found that the well-known Wi-Fi-enabled Wemo Insight Smart Plug has critical security vulnerability due to Universal Plug and Play (UPnP) protocol library it uses, which, due to design flaws, enable attackers to execute remote codes on this smart plug. Note that this attack is not just limited to disturbing smart plug's normal operations such as shutting it down unexpectedly, but could also use the smart plug as an entry point for a larger attack in the network. Further,~\cite{guoxi1} showed the usage of a compromised WeMo switch as a middleman to launch attacks against a TCL smart TV.

The privacy vulnerabilities introduced by smart home devices are even more challenging. IoT devices capture, store, share, and (depending upon the underlying computational architecture) outsource personal data that can lead to inferences about individual's habits, behavior, family dynamics etc. Challenges arise since privacy leakage can occur through direct data leakage, as well as, through inferences based on device actuations, interactions, and schedules. For instance, the timing of the actuation of a coffee machine, if leaked can allow an adversary to determine when a family wakes up. Likewise, locking and unlocking schedule of door locks can enable leakage of the time when no one is at home, etc. While privacy challenges from data leakage can be prevented by encrypting device data and network payloads, inferences about device actuation and schedule are significantly more complex to hide due to leakage from network traffic patterns at the channel level, at the hub/router level or at the cloud level.

\begin{figure}[h]
	\begin{center}
		\includegraphics[scale=0.5]{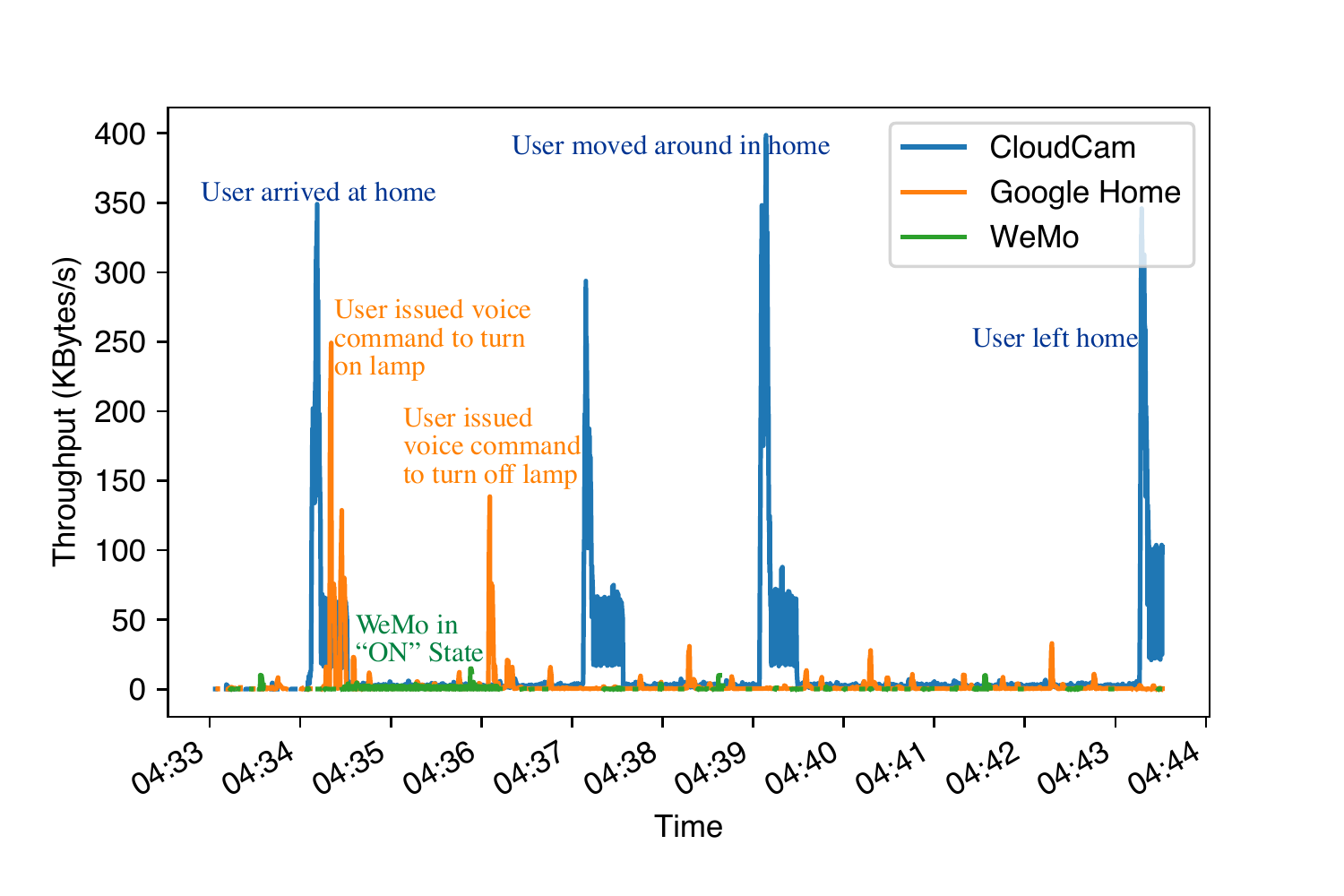}
	\end{center}
\caption{Channel activity for home devices.}
	\scriptsize{The figure above shows the channel activity for three home devices: CloudCam, Google Home, and Belkin WeMo. The CloudCam shows a peak in the channel activity (up to 400 KB/s traffic rate) as the user enters the home, moves inside the home or exits the home. The Google Home shows a peak in the channel activity (up to 250 KB/s traffic rate) whenever a user initiated a voice command for the light bulbs to turn on/off. Similarly, a bi-state WeMo switch peaks during the on state and creates a channel activity lesser than 20 KB/s.}
	\label{fig:peak}
\end{figure}

For example, Figure~\ref{fig:peak} shows the channel traffic generated by three different devices. The figure clearly shows that each device generates a very distinct traffic pattern and, the adversary, having access to the channel traffic can figure out which device is activated leading to potential inferences about user's personal habits. Note that such an inference, since it is independent of the actual network payload, is not prevented by encryption.

Inferences from monitoring channel traffic can also arise due to the characteristics of the current network protocols. For instance, in
the widely used 802.11 Wi-Fi protocol, while a message payload is encrypted in a password-protected network, the MAC addresses of both the sender and receiver are in cleartext. This is to prevent requiring every potential device on the network to have to decrypt a message just to determine if the message is intended for the device. The leakage of the sender/receiver MAC address, coupled with the fact that manufacturer's information is commonly encoded in plain-text device identifier, can lead to leakage of the identity of the device from the network traffic, which, in turn, can lead to an attack on user's privacy. For example, the MAC address of Amazon CloudCam security camera used in our experiment to transmit video footage over Wi-Fi is F0-81-73-23-CC-75. The first 3 bytes of such a MAC address (e.g., F0-81-73) can be searched in the publicly accessible IEEE Organizationally Unique Identifier (OUI)~\cite{2guoxi} dataset to find the vendor related information (e.g., Amazon device). Furthermore, by monitoring the device's traffic patterns and the fact that Amazon only manufactures a limited range of devices (e.g., Kindle, CloudCam, Echo, etc.), it is easier to infer the device type by merely overhearing the traffic.

In this paper, we study privacy leakage that may occur from device activity and network channel traffic analysis and develop protocols that can be used to prevent such leakages. We focus, in particular, on device workflows that are common in smart homes. By a device workflow, we refer to a coordinated sequence of device actuations. Device workflows may arise in a {\em triggered} (or {\em synchronized}) manner or in a {\em scheduled} manner. The synchronized workflows arise as a result of one device resulting in an actuation of the other. For instance, sensors determining occupancy change in a part of the building may result in HVAC controls /AC vents to be redirected to the occupied areas and to close other vents that cover areas with no occupancy. Likewise, light intensity sensors coupled with thermal sensors may detect the amount of sunlight entering the room and accordingly lower/raise the sunshades based on the homeowner's preference. A scheduled workflow, on the other hand, is scheduled actuation of a set of devices that occur at specific time intervals of each other based on a schedule. For instance, switching on a coffee machine at a specific time in the morning followed by warming of the car seats a given time interval following that, and then opening/closing of door locks following the actuation of the car seat warmer might be on a schedule. A more elaborate example of a scheduled workflow could be a homeowner's routine related to returning. A homeowner may schedule the smart car to self-drive to home to initiate the workflow. Fifteen minutes after the start of the workflow, the heating/cooling system may start off to ensure that the home is at the comfortable temperature on arrival. Likewise, half an hour after the trip starts the oven may be set to a pre-heat and the laundry machine turns on if the load is detected.

We focus on the scheduled workflows in this paper since such workflows require hiding the identity of the devices being actuated but also their schedule. As will become clear, mechanisms to prevent leakage for scheduled workflows will also prevent leakage from triggered workflows. Furthermore, the scheduled activities can lead not just to adversary learning user's past behavior but also their future activities which can lead to more severe consequences.


\smallskip
\noindent\textbf{The problem.} This paper deals with a problem of avoiding inference attacks on the scheduled workflows in a home network. The workflows can be identified through {\em coupling} between the {\em channel} and {\em device activity}. Basically, there are two crucial concepts that are subject to privacy violations: workflow (i.e., the specific order of device actuation) and workflow execution (i.e., duration in which the devices coordinate, and the resulting device actuations unfold). Our problem statement considers hiding both the workflow and the execution of the workflow.

The privacy violations can occur as a result of two threats: first, overhearing the channel activity as a means to infer device activity pattern, second, accessing the device temporarily and be able to analyze the state of workflow execution. In the latter case, the adversary can read the sent/received messages or the internal state of the device. Both of the above threats may assist the adversary to predict users' activities, such as presence/absence, arrival/departure, and localization etc.

\smallskip
\noindent\textbf{Contributions.}
Our contributions are twofold:
\begin{enumerate}[noitemsep,nolistsep,leftmargin=0.1in]
  \item A new architecture for in-home communication among the devices and the hub through passing a token carrying commands. The token passing communication model decouples the channel and device activities so that the devices interact with the hub and the other devices without revealing the communication pattern. In addition, this architecture is also useful for secure data upload from devices to the hub, while also hiding the device footprints that has generated the data.

  \item We provide an owner-defined pre-scheduling mechanism for all devices that are connected with the hub in a pre-defined topology. The proposed approach uses a single message transmission for all $N$ devices while ensuring that the in-home communication remains peakless. The scheduling mechanism is secure against a computationally unbounded adversary and, also, verifies the delay between each device actuation.
\end{enumerate}

\noindent\textbf{Outline.} The paper proceeds as follows: Section~\ref{sec:Preliminaries} provides the system setting, the adversarial model, security goals, and design requirements. Section~\ref{sec:decoup} provides our proposed scheduling algorithm for home networks. Section~\ref{sec:analysis} provides proofs of security and privacy. Finally, Section~\ref{sec:Experimental Evaluation} provides an experimental evaluation of the approach. All notations are given in Table~\ref{table:notations}.

\section{Preliminaries}
\label{sec:Preliminaries}
This section presents the system model, the adversarial model, inference attacks on the user privacy, an overview of our proposed approach to prevent inference attacks, design requirements, and building blocks of the proposed algorithms.

\begin{table}[t]
	\begin{center}
		\begin{tabular}{l l l l}
			\hline
			{\scriptsize Notations} & {\scriptsize Meaning} & {\scriptsize Notations} & {\scriptsize Meaning} \\ \hline
			
			{\scriptsize $O$} & {\scriptsize A homeowner} & {\scriptsize $H$} & {\scriptsize Hub} \\
			
			{\scriptsize $O_{\mathit{id}}$} & {\scriptsize Owner's identity} & {\scriptsize ${H}_{\mathit{id}}$} & {\scriptsize Hub identity} \\ 
			
			{\scriptsize $D$} & {\scriptsize Device} & {\scriptsize $D_{id}$} & {\scriptsize Device identity} \\ 
			
			{\scriptsize $O_{\mathit{PK}}$} & {\scriptsize Owner public key} & {\scriptsize $O_{\mathit{SK}}$} & {\scriptsize Owner secret key} \\ 
			
			{\scriptsize $H_{\mathit{PK}}$} & {\scriptsize hub public key} & {\scriptsize $H_{\mathit{SK}}$} & {\scriptsize hub secret key} \\ 
			
			{\scriptsize ${D}_{\mathit{PK}}$} & {\scriptsize device public key} & {\scriptsize ${D}_{\mathit{SK}}$} & {\scriptsize device secret key} \\ 
			
			{\scriptsize $c_l$} & {\scriptsize partially ordered $l$ commands} & {\scriptsize $m^i$} & {\scriptsize puzzle message for $i$th device} \\ 
			
			{\scriptsize $n$} & {\scriptsize modulus} & {\scriptsize $a$} & {\scriptsize random chosen integer} \\ 
			
			{\scriptsize $\hat{t}$} & {\scriptsize time complexity of puzzle} & {\scriptsize $t^{\prime}$} & {\scriptsize time to decrypt command} \\ 
			
			{\scriptsize $S$} & {\scriptsize capacity of puzzle solver} & {\scriptsize $S^{\prime}$} & {\scriptsize enhanced capacity}  \\ 
			
			{\scriptsize $\mathit{Sign}(O_{\mathit{SK}})$} & {\scriptsize Signature using secret-key $O_{\mathit{SK}}$} & {\scriptsize $N$} & {\scriptsize number of devices} \\ 
			
			{\scriptsize $E_k$} & {\scriptsize encrypted key $k$} & {\scriptsize $E_z$} & {\scriptsize encrypted command $z$} \\ 
			
			{\scriptsize $k_s$} & {\scriptsize static key} & {\scriptsize $^nP_n$} & {\scriptsize $n$ permutations} \\ 
			
			{\scriptsize $\phi{(n)}$} & {\scriptsize Euler's totient on $n$} & {\scriptsize $b_o$} & {\scriptsize overwritten data bits} \\ 
			
			{\scriptsize $b_g$} & {\scriptsize device generated data bits} & {\scriptsize $b_r$} & {\scriptsize random data bits} \\ 
			
			{\scriptsize $p$} & {\scriptsize large prime number} & {\scriptsize $q$} & {\scriptsize second large prime number} \\ 
			
			{\scriptsize $t_{\mathit{val}}$} & {\scriptsize command validity time} & {\scriptsize $t_{\mathit{cur}}$} & {\scriptsize current time} \\ 
			
			{\scriptsize $t^i_{\mathit{rcv}}$} & {\scriptsize token receiving time} & {\scriptsize $t^i_{\mathit{fwd}}$} & {\scriptsize token forwarding time} \\ 
			
			{\scriptsize $t^H_{\mathit{beg}}$} & {\scriptsize token round beginning time at hub} & {\scriptsize $t^H_{\mathit{end}}$} & {\scriptsize token round ending time at hub} \\ 
			
			{\scriptsize $t_{\mathit{diff}}$} & {\scriptsize allowed clock drift time} & {\scriptsize $t_{\mathit{com}}$} & {\scriptsize total computation time} \\ 
			
			{\scriptsize $t^{\mathcal{A}}_{\mathit{com}}$} & {\scriptsize puzzle computation time by adversary} & {\scriptsize $\epsilon$} & {\scriptsize negligibly small value} \\ 
			
			{\scriptsize $\mathcal{A}$} & {\scriptsize adversary} & {\scriptsize $\hat{a}$} & {\scriptsize malicious commands} \\ 
			
			{\scriptsize $\mathcal{T}$} & {\scriptsize token} & {\scriptsize $\mathcal{H}$} & {\scriptsize One-way hash function} \\ 
			
			{\scriptsize $\mathit{Data \: field}$} & {\scriptsize data upload field} & {\scriptsize $b_{\mathit{toggle}}$} & {\scriptsize toggle bit string field} \\
			
			{\scriptsize $\mathcal{R}$} & {\scriptsize partial order} & {\scriptsize $\mathcal{E}^\prime$} & {\scriptsize partially ordered set} \\ \hline
		\end{tabular}
		\caption{Notations}
		\label{table:notations}
\BBB\BBB\BB
	\end{center}
\end{table}

\subsection{The Model}\label{subsec:mode}
\noindent\textit{Network assumptions}. We consider a homeowner, $O$, who owns a collection of $N$ ($D^1,D^2,\ldots,D^N$) heterogenous smart home devices that provide different functionalities to the owner. Each device $D^i$ has a unique identity, denoted by $D^i_{\mathit{id}}$. All devices might possess heterogeneous hardware/software underneath, and be located on different spatial (devices that are not in the line-of-sight) dimensions. These ad-hoc devices can shift in space in the smart home, and hence, might have a different set of peer devices at different time intervals. We assume that each device possesses a read-only hardware clock, and due to the ad-hoc nature of devices, we assume a clock drift within the bound $t_{\mathit{diff}}$ such that two clocks cannot differ beyond the $t_{\mathit{diff}}$ amount of time.

The owner initializes the devices and a controlling hub, $H$, using proper security mechanisms. We will list our assumptions about the underlying security mechanism below. In our model, the network is configured as a ring topology, which poses an ordering among devices, unlike the model that the current smart home devices use, where the owner communicates directly to the desired device via the hub. This ring topology could be built directly among devices and hub, if the communication protocols they use have P2P communication capability, like Zig-Bee, BLE, and Wi-Fi. Alternatively, it could be built as an overlay on top of a star topology network, like Wi-Fi infrastructure mode. In this case, if a device tries to forward the message to the next device in the proposed ring topology, then it needs to first send the message to the hub, and the hub, then, directly forwards the message to its next device. Note that we do not discuss a failure-resilient ring topology and existing fault-tolerant schemes can be leveraged here.

The owner sends workflows to the home devices through the hub. After receiving a workflow, the device gets actuated, stores its corresponding command, and forwards the workflow received from the previous device to the next device in the topology. After executing the command, the device may generate the data (for example, Nest camera starts recording and sends data whenever motion is detected). We use the ring topology to send this data to the hub that may be stored at the hub or may be transmitted to the cloud.\footnote{Recall that we are not dealing with how the data will be transmitted from the hub
to the cloud without revealing anything. Our solution hides any activity within the home, i.e., how the data will be transmitted by the device to the hub without any privacy violations.} In this paper, we use the words `command execution' and `workflow execution' interchangeably.

\noindent\emph{Token.} In our ring topology, we circulate a token that has three fields: (\textit{i}) command field, which carries a computational puzzle and the workflow, (\textit{ii}) data field, which carries the data generated by devices to deliver to the hub, and (\textit{iii}) toggle bit string, which is used to indicate which device has generated the data to the hub. Details of the token are given in Section 3.

\noindent\textit{Security assumptions}. Each device, hub, and homeowner possess the corresponding signing key-pair, i.e., ($D^i_{\mathit{SK}},D^i_{\mathit{PK}}$), ($H_{\mathit{SK}},H_{\mathit{PK}}$), and ($O_{\mathit{SK}},O_{\mathit{PK}}$), respectively. We do not assume an arbitrary behavior from the owner or the hub. The \emph{homeowner and hub} mutually verify the identities of each other through digital signatures in order to build the trust between them. Therefore, the hub and the owner trust each other, and an adversary cannot compromise either the homeowner or the hub. Further, \emph{the hub and devices} build their own trust that is also based on the knowledge of a certified public-private key-pair of home devices.

\subsection{Adversarial Model}
\label{subsec:Adversarial Model}
The devices execute user-defined commands or workflows, as mentioned previously. The adversary wishes to learn the commands or workflows and the execution of workflows based on the encrypted network traffic to infer the user privacy. This type of adversary is similar to the adversary considered in~\cite{watson,thir}. Thus, the adversary has access to the secure (encrypted) messages flowing among the devices and the hub/homeowner. Further, we assume that the adversary knows the number of smart devices in the home. Based on this information, the adversary aims to learn: (\textit{i}) the device activities, and (\textit{ii}) coupling between the channel and the device activities. However, the adversary cannot inject any fabricated messages over the channel to assess the state of the devices.

Further, we assume that the adversary may gain a short-term physical access\footnote{For example, an inspection authority has to visit the home for a periodic inspection in the absence of the homeowner. In this case, the inspection might be related to any leakage detection, maintenance issues, insurance issues, etc. This short visit to the home for an inspection allows them to monitor and check home devices as well. However, based on our proposed solution, those inspecting authorities would not be able to analyze the current state of the devices or the activity pattern of the devices in near future.} to the device, and hence, can retrieve the device state or messages. The objective behind gaining a short-term access to any device is to predict the future workflows of devices to infer the user activity. Our objective is to prevent the adversary to know (\textit{i}) which device has received the messages at which time, (\textit{ii}) when would a device execute the command, and (\textit{iii}) which devices have executed the command at which time\footnote{The command execution could enable the device to produce visible or auditory cues such as blinking lights or machine being activated which, in turn, may leak the state of the device. Such inferences from physical cues are outside the scope of the paper. We assume that the adversary does not have access to such device data.}.
	

\noindent\textbf{Adversarial view and inference attacks.} When the user wishes to execute any command at a smart home device, an adversary knows which device received the message from the user at what time due to the network traffic generated by the user. Note that this information is revealed, because in network protocols such as 802.11 MAC addresses or the device identifiers are transmitted in cleartext, and only the payload is encrypted, as mentioned in Section~\ref{sec:introduction}. Further, the device may also produce some data in response to the requested message, and this also reveals to the adversary which device has generated the data at what time. We refer such information as the adversarial view, denoted by $\mathit{AV}$:
$\mathit{AV} = \mathit{In}_c \cup \mathit{Op}_d$, where $\mathit{In}_c$ refers to the command given to the device at some time and $\mathit{Op}_d$ refers to the data generated by the device after executing the command. 

\begin{table}[h]
\BB
	\scriptsize
	\centering
	\begin{tabular}{|l|l|l|}
		\hline
		Users command & \multicolumn{2}{|c|}{Adversarial view}           \\ \hline
		~                          & $In_c$ & $Op_d$ \\ \hline
		For $D^1$ & $\mathit{E(c_1)}, D^1, t_1$        & No        \\ \hline
		For $D^2$ & $\mathit{E(c_2)}, D^2, t_2$        & $\mathit{E(d_2)}, D^2, t_3$        \\ \hline
	\end{tabular}
	\caption{Adversarial view.}
	\label{tab:Adversarial view}
\BBB\BBB\BB
\end{table}

For example, consider that there are two devices, say $D^1$ and $D^2$, in the home. In Table~\ref{tab:Adversarial view}, the first row shows that the user transmits a command to the device $D^1$. Though the device $D^1$ receives this encrypted command, denoted by $E(c_1)$, the adversary knows that a command is received at time $t_1$ by the device $D^1$ and the device $D^1$ has not generated any data in response the command. The second row shows that the adversary knows the device $D^2$ receives an encrypted command $E(c_2)$ at time $t_2$ and generated encrypted data $E(d_2)$ at time $t_3$. Hence, simply based on the above characteristic of the arrival of a message and generation of data, the adversary can determine which device was actuated.

\subsection{Preventing Inference Attacks: overview of our Approach}
\label{subsec:Preventing Inference Attacks}
In order to prevent inference attacks, we develop an approach that decouples the device and channel activities. In short, the approach provides an ability to pre-schedule a set of commands for home devices, where the homeowner defines: (\textit{a}) what should be the workflow/schedule of home devices, and (\textit{b}) when should the devices execute a workflow. Informally, the proposed approach works as follows:


\begin{enumerate}[noitemsep,nolistsep,leftmargin=0.01in]
\item The owner invokes the hub by sending an encrypted schedule or workflow of the devices. Here, the hub authenticates the owner to validate the encrypted schedule. This step provides a guarantee that no channel spoofing or message re-transmission have occurred.
	
\item After a successful authentication phase, the hub creates a token ($\mathcal{T}$) to circulate the schedule to be executed by devices. This token rotates continuously in the topology. Note that in the token, device identifiers or MAC addresses are also encrypted. Further note that whenever the user wishes to transmit a schedule to devices, the immediate next round of token originated by the hub carries the encrypted schedule, and after that, the encrypted schedule is replaced by a random message to maintain the token size constant.
	
	\item On receiving the token, each device retrieves the encrypted schedule that carries device-specific commands. Each device must complete a computation task before executing the original command. This computation task is referred to as a puzzle throughout the paper. Then, the device must check the puzzle validity time $t_{\mathit{val}}$, by using the current time $t_{\mathit{cur}}$, and the allowed clock difference $t_{\mathit{diff}}$. If the timer has not expired yet, the device decrypts the message, executes the puzzle to retrieve the real command to be executed.
	
	\item As soon as a device finishes the command execution, it may generate data to be uploaded at the hub (as mentioned in Section~\ref{subsec:mode}). Now to hide which device has generated the data, each device follows a request-based approach, where a device $D^i$ flips the $i^{\mathit{th}}$ bit inside one of the fields of the token to indicate the need to upload freshly generated data in an anonymous manner. As a result of $i^{\mathit{th}}$ bit flipping, the hub knows that the device $D^i$ has requested the data upload, and in the next token cycle, the device $D^i$ appends the data inside a dedicated field of the token.
\end{enumerate}
Note that since a constant size token flows regularly in the topology of the home devices, an adversary observing the network traffic cannot distinguish which device has received a command at which time (due to step 2) and which devices have generated the data (due to step 4). Hence, based on this approach, the adversarial view for each round of token has the same information, which prevents inference attacks based on the devices and channel activities.

\subsection{Security Goals}
This section describes the security properties for preventing any inferences about the workflow and their executions from the adversary. Let us assume that an adversary knows some auxiliary information about the devices and the topology such as the number of devices and the types of devices. However, this auxiliary information does not increase the probabilistic advantage that an adversary gains over any instance of the protocol. In particular, an adversary cannot reveal the workflow or the execution time of the workflow, i.e., which devices execute the command or when does a device execute the command. The probabilistic advantage of an adversary, denoted by $\mathit{Adv(\mathcal{A})}$, is derived through the security properties given below.

\smallskip\noindent\textbf{Authentication} is required during the workflow release from the homeowner to the hub. This would require a mutual authentication between the (mobile device held by the) homeowner (to dispatch the workflow) and the hub (to circulate the workflow anonymously). Note that establishing the shared secret between the homeowner and the hub is a one-time process, which is carried each time the homeowner invokes a new workflow. Here, the homeowner produces a signature, say $\mathit{Sign}$, on the ordered commands ($c_l$) in any workflow by using its secret-key $O_{SK}$. Thus, the hub must reject any other messages signed by a different key, say $O_{{SK}^\prime}$.

\centerline{$Pr[(O_{SK},c_l)\rightarrow \mathit{Sign}]\geq 1-\epsilon$}

Note that $\epsilon$ is negligibly small and an adversary cannot produce a verifiable signature $\mathit{Sign}$ on $c_l$ by using $O_{{SK}^\prime}$ instead of $O_{SK}$.

\smallskip\noindent\textbf{Anonymity} is required during the consistent circulation of the encrypted commands such that (\textit{i}) no channel activity can be mapped to a device activity, and (\textit{ii}) no inference on device activity can be mapped to the device generated data, i.e., which device is sending data at a specific time. As shown below, the probability of distinguishing two different tokens ($\mathcal{T},\mathcal{T}^\prime$) each carrying different messages ($m^i,m^j$) for different devices ($i,j$) is negligible.

\centerline{$Pr[\mathcal{T}(m^i)]-Pr[\mathcal{T}^\prime(m^j)]< \epsilon$}

Similarly, the probability of distinguishing a token $\mathcal{T}$ carrying the random data $b_r$ or carrying the overwritten data $b_o$, is negligibly small. Therefore, a token carrying the random data inside the data field and another token carrying the overwritten data (after data generation) inside the data field are indistinguishable, hence, solely based on the token data field no inferences can be derived.

\centerline{$Pr[\mathcal{T}(b_r)]-Pr[\mathcal{T}(b_o)]< \epsilon$}

\smallskip\noindent\textbf{Verifiable delay} An adversary cannot infer the information about the device execution ahead of time. Let $\hat{t}_i$ be the time a device would execute even when the adversary has temporary access to the device.

\centerline{$Pr[t^{\mathcal{A}}_{\mathit{com}}|\mathit{state}] \approxeq Pr[t^{\mathcal{A}}_{\mathit{com}}]$}

The probability of an adversary finishing the computation task earlier, when it gains temporary access to the device state, is approximately same as when the adversary does not have access to the device state. In addition, an adversary cannot outpace a device that requires $\hat{t}^i$ time to complete the computation task, i.e., $Pr[t^{\mathcal{A}}_{\mathit{com}}<\hat{t}_i]< \epsilon$.


An adversary gaining access to the device cannot retrieve the information required for device actuation, i.e., the computational task to be executed prior to its actuation.
We have described a game-theoretic approach in Section~\ref{sec:analysis} that shows the overall probabilistic advantage of an adversary is negligibly small.





\B
\subsection{Challenges and Solutions}
\label{subsec:Challenges and Solutions}\B
Implementing reliable ordering for device actuation is to provide a secure and self-executing state of devices at a pre-defined time for protecting the owner privacy is deceptively non-trivial. Below, we discuss the challenges we encountered and describe how we addressed them.

\smallskip\noindent\textbf{C1. Anonymous trigger from the hub to devices.} As mentioned before, we need to mask the channel activities, device activities, and the coupling between both the channel as well as the devices. Note that in this context, the encryption techniques merely hide the meaning of the message across the channel, not the fact that to which device this message belongs to, and hence, it reveals the user activity. In addition, the solutions-based on {\em traffic shaping}\footnote{The traffic shaping solution keeps the constant traffic rate based on a threshold, such that any excessive traffic above the threshold is delayed through a buffer and below the threshold requires additional dummy packets.}, which incur excessive communication and latency overhead, also fail to decouple the device to channel activities.

\fbox{\begin{minipage}{3.3em}
		\textit{Solution.}
\end{minipage}}
To address this challenge, the distribution of user-commands to each device in the home network is based on a pre-defined topology, e.g., ring, where a token rotates continuously within the one directional (1-D) ring topology.\footnote{In order to leverage a continuous channel-activity as a means to hide the actual channel-activity, the ring topology is efficient as compared to star alignment. In addition, one can also use the mesh-topology with anonymous-routing that requires asymmetric-key cryptography overheads at each relay-node while sending token between a source and a destination.} Thereby, channel activity remains consistent and independent of the devices actuated as a result of the workflow. In our context, the token ($\mathcal{T}$) has three fields: (\textit{i}) command field, which contains the encrypted commands corresponding to each device, (\textit{ii}) data field, which contains the device generated data, and, (\textit{iii}) a toggle bit string field, which contains a $N$ bit string, where each bit denotes a unique device in the topology. This toggle bit string is used to indicate that a device is interested in uploading the freshly generated data during upcoming token arrival at the device. Since the order of device actuation reveals crucial information about the user activity inside the home, our token-based solution guarantees a secure ordering among devices while executing the commands. In addition, the device actuation is controlled in a manner that a recipient device itself cannot pre-decode and/or pre-pone the command execution.


\smallskip\noindent\textbf{C2. Command execution and verifiable ordering.} After masking the channel activity through a constantly rotating token, the next step is to have a verifiable ordering of command execution at each device. Each device receives a command through the token. Now, these devices can decrypt and execute the command immediately. However, it again enables the channel to device activity mapping. Therefore, the next challenge is to insert an artificial delay between a device receiving the commands, and then, executing the commands at an appropriate time, without relating to any specific channel activity. The artificial delay enables a correct execution order at each comparable or non-comparable device\footnote{The comparable-devices are those that are defined under certain relative order in a workflow. The incomparable-devices are independent and are not restricted under any relative order with respect to other devices. }.

\fbox{\begin{minipage}{3.3em}
		\textit{Solution.}
\end{minipage}} The protocol message from the homeowner to hub includes the ordered commands ($c_{l}$) that pass through a device $D^i$ to another device $D^{i+1}$ using the anonymous token circulation. The token is encrypted using a shared symmetric key between the hub and the devices in the topology. 
The recipient device $D^i$ retrieves the encrypted command ($m^i$) from the command field of the token and begins with a puzzle computation. Note that the devices do not execute the commands immediately after receiving (time-locked) commands. In particular, these devices wait for a pre-defined amount of time before executing the command, such that neither the artificial delay at each device can be known by the adversary in advance, nor the devices can control this artificial delay to postpone or prepone the scheduled commands. However, this waiting period is not idle, and the devices resume on some computational task.

\smallskip\noindent\textbf{C3. Anonymous response from devices to the hub.} Once the devices have received the commands in an anonymous manner, they may generate some data as a result of command execution. However, uploading this data immediately would reveal the device activity patterns. Clearly, this periodic channel activity (in the form of traffic) relates to a specific device that had executed the command recently. Therefore, the upstream data upload on the hub should be anonymous too.

\fbox{\begin{minipage}{3.3em}
		\textit{Solution.}
\end{minipage}} In our scheme, each device is required to send a request for data upload to the hub using the rotating token, which contains a toggle bit string field of size $N$ bits, where each bit represents a unique device in the topology. The $i^{\mathit{th}}$ toggled bit indicates that the device $D^i$ has generated the data and is ready to anonymously transmit the data by using the data field of the token. The data field is used to carry the device generated data without revealing the data and the sender of the data. The data field contains random data as long as there is no request from the devices to send the data to hub. The details of data upload phase from device to hub will be clear in step 4, given in Section~\ref{sec:decoup}.




\B
\subsection{Building Blocks} This section provides a brief overview of basic building blocks used to develop the proposed solution as detailed in Section~\ref{sec:decoup}.

\noindent\textbf{RSA puzzles.} The verifiable delay regarding device actuation is based on cryptographic RSA puzzles~\cite{rlock}. These time-bounded puzzles are useful for the applications that require security against the hardware parallelization attacks (i.e., bypassing a security solution by running a mathematical problem on different hardware in parallel) through Application Specific Integrated Circuits (ASIC). Accordingly, the puzzle solution is based on inherently sequential operations such as {\em modular exponentiation}. Let us assume that $t^{\prime}$ be the time to release an encrypted message. Also, a device is capable of computing $S$ number of square operations modulo $n$ per second. Thus, the puzzle requires sequenced exponentiations of ($a^{2^{\hat{t}}}$ {\em mod} $n$) where $n=pq$ is publicly known RSA modulus and $\hat{t}$, $p$, $q$ and $\phi(n)=(p-1)(q-1)$ remains secret. In particular, $\hat{t}=St^{\prime}$ denotes the difficulty level of the puzzle for a specific device. Therefore, the computation of this modular exponentiation operation requires either the inherently sequential execution of these operations {\em or} to solve the integer factorization problem.

In the proposed scheme, the RSA puzzles enable an artificial yet verifiable delay with respect to command execution. To compensate this delay an adversary must know the private key of a device and then invest the same time as the victim device was supposed to invest in, for the puzzle computation. In particular, the adversary can always lengthen the delay (which is easily detectable during the puzzle validity check), but cannot shrink the delay due to inherently sequential operations.

\noindent\textbf{Order-preserving bijection.} The order-preserving bijection guarantees an instance of the totally ordered set elements as derived from the partially ordered set of the same elements. In particular, the bijection provides a unique sequence of the totally ordered set elements. Let us assume there is a partial order relation $\mathcal{R}=\{{\leq}\}$ on set elements $\mathcal{E}=(e^1,e^2,e^3,e^4)$ such that $\mathcal{R}$ is: reflexive, i.e., $e^i\mathcal{R}e^i$; antisymmetric, i.e., if $e^i\mathcal{R}e^j$ and $e^j\mathcal{R}e^i$ then $e^i=e^j$; and transitive, i.e., if $e^i\mathcal{R}e^j$ and $e^j\mathcal{R}e^k$ then $e^i\mathcal{R}e^k$. In addition, the partially ordered set $\mathcal{E}^\prime=((e^1,e^2),(e^3,e^4))$ under relation $\mathcal{R}=\{\leq\}$ has a unique minimal and maximal element. Therefore, an order-preserving bijection generates a linear ordering of elements in set $\mathcal{E}^\prime$. Essentially, this linear extension generates a permutation order of the elements in a given partially ordered set $\mathcal{E}^\prime$. All of those permutation sequences in which $e^i$ appears before $e^j$ given that $(e^i\leq e^j) \: \in\mathcal{E}^\prime$ are a valid candidate as per any totally ordered set element sequence. The reliable ordering of device actuation is guaranteed through the linear extension of the owner-defined schedule even when the application host is unavailable.

\section{Decoupling Channel Activity from Device Activity}
\label{sec:decoup}
This section provides the details of our proposed protocol for decoupling channel activities from device activities. First, we illustrate the device-to-device interaction through an example as below:


\begin{figure}[t]
	\BBB
	\begin{center}
		\includegraphics[scale=0.35]{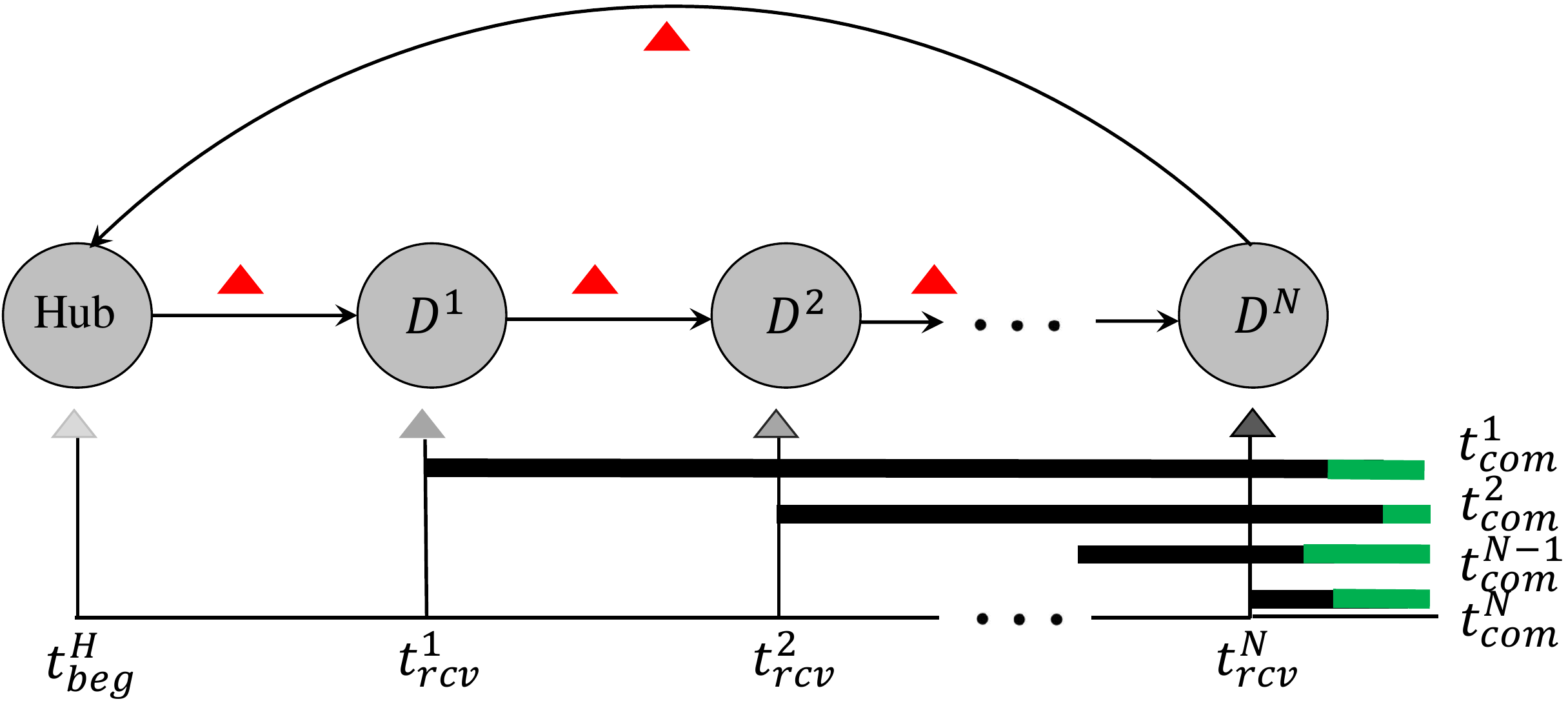}
	\end{center}
	\BBB
	\caption{Device ordering on the timeline.}
	\BBB
	\label{fig:problem statement}
	\BBB
\end{figure}

\subsection{Example}
Consider $N$ number of devices ($D^1,D^2, \ldots,D^N$) that are connected through a hub ($H$), as shown in Figure~\ref{fig:problem statement}. Let $s_{\mathit{on}}^i$ (and $s_{\mathit{off}}^i$) be the on (and off) state of a device $i$. The owner $O$ can create a partial ordering for devices such as
$\langle(D^1,D^2,D^3),$ $(D^4,D^5),$ $\ldots,$ $(D^{N-1},D^N)\rangle$ based on their states, e.g., $\langle(s^1_{\mathit{on}},s^2_{\mathit{off}},s^3_{\mathit{off}}),$ $(s^4_{\mathit{off}},s^5_{\mathit{on}}),$ $\ldots,$ $(s^{N-1}_{\mathit{off}},s^N_{\mathit{on}})\rangle$
that shows the device $D^1$ must change its state to on, i.e., $s^1_{\mathit{on}}$, before the devices $D^2$ and $D^3$ change states to off, i.e., $s^2_{\mathit{off}}$ and $s^3_{\mathit{off}}$. Similarly, the device $D^4$ must change its state to off, i.e., $s^4_{\mathit{off}}$, before the device $D^5$ changes its state to on, i.e., $s^5_{\mathit{on}}$.\footnote{For the sake of simplicity this example includes the ordering between the devices and the corresponding states. However, throughout the paper, our focus is to order time intervals for devices' actuation.} After creating the partial order of devices, the owner sends a message to the hub that sends the message (shown in red color) to one of the devices, as shown $D^1$ in Figure~\ref{fig:problem statement}. We refer to the message from the hub to devices as a token. Each device $i$ receives the token at time $t^i_{\mathit{rcv}}$, forwards the token, and begins computation at time $t^i_{\mathit{com}}$. The bottom part shows when each device receives the token in a sequence as ($D^1,D^2,\dots,D^{N-1},D^N$). Note that due to user-defined partial order of device actuation $\langle(D^1,D^2,D^3), (D^4,D^5),\ldots, (D^{\mathit{N-1}},D^N)\rangle$, the devices across the partial orders are mutually incomparable\footnote{In case, the user finds a change in his/her schedule, then another remote command can overwrite the previous commands and the schedule workflow, correspondingly.}. In Figure~\ref{fig:problem statement}, thick black line (for each device) shows that the device is having the token and waiting for the predefined time (given in the token) for its activation, and the green line shows when the devices start working.

\subsection{Verifiable Ordering Protocol}
The Section formally defines the verifiable ordering protocol and a detailed description for each step as below.

\begin{definition}[Verifiable Ordering Protocol]
The verifiable device ordering protocol is a tuple $(\mathit{param},\mathit{puzgen},P_D^i,P_O)$ of four polynomial-time algorithms such that:
\end{definition}
\begin{itemize}[noitemsep,nolistsep,leftmargin=0.05cm]
\item \emph{Public parameter generator}. $\mathit{param}(1^{\lambda}) \rightarrow (n, a)$. $\mathit{param}(1^{\lambda})$ initializes the prime integer factors $n=pq$ (where $p$ and $q$ are two large prime numbers) and random value $a$ for the puzzle creation.

\item \emph{Puzzle generator}. $\mathit{puzgen}(n, a, \hat{t}, E_z, E_k) \rightarrow \mathcal{P}$. $\mathit{puzgen}(n, a, \hat{t}, E_z, E_k)$ selects the input values as target time for commands execution $\hat{t}$, encrypted command $E_z(z,k)$, encrypted key $E_k(a,\hat{t},n,k)$ and outputs a puzzle $\mathcal{P}$ for each device.

\item Follower ${D^i}(\mathcal{P}^i, SK) \rightarrow z(\hat{t}_i, t^i_{\mathit{com}}]$ completes the puzzle $\mathcal{P}^i$ and executes the command within the half-open interval, i.e., no earlier than $\hat{t}_i$ but earlier than or at $t^i_{\mathit{com}}$.

\item Owner $O(\mathcal{P}, \phi{(n)}, PK^i) \rightarrow \mathit{accept}(t^i_{\mathit{com}}\geq \hat{t}_i)$ accepts the timely command execution at each device using $\phi{(n)}$.

\end{itemize}

\noindent\textit{Setup and key distribution.} The manufacturing authority initializes a unique identity for the owner, hub, and, home devices by using the secure identity distribution function, say $\texttt{Init}(1^\lambda)\rightarrow \mathit{identity}$, where $1^\lambda$ is the security parameter that generates a unique $\mathit{identity}$ for each entity. The certificate authority verifies that each of these devices knows the private key paired to the public key proposed for certification as: the homeowner ($O_{\mathit{SK}},O_{\mathit{PK}}$), the hub ($H_{\mathit{SK}}, H_{\mathit{PK}}$), and $i^{\mathit{th}}$ home device ($D^i_{\mathit{SK}},D^i_{\mathit{PK}}$), possesses a valid key pair.



\LinesNotNumbered
\begin{algorithm}[t]
	\DontPrintSemicolon
	\textbf{Inputs:} set of $l\in N$ devices ($D^i$), public keys ($D^i_{\mathit{PK}}$)
	
	{\bf Variables:} $\mathcal{P}$ a puzzle, $z$ a command.
	
	\nl{\bf Function $\mathit{create} \: (c_l,D^N)$} 
	
	\nl \Begin{
		\nl \For{$\forall (i,j) \: \exists (D^i,D^j) \: \in H$}{
			\nl $\mathit{schedule} \: ((D^i,D^{\mathit{i+1}}),(D^j,D^{\mathit{j+1}}))$
			
			\nl \For{$\forall (i,i+1)$}{      
				
				\nl $z=(s_{\mathit{on}} \vee s_{\mathit{off}}) \wedge (\hat{t}_i \leq \hat{t}_{i+1})$
				
				\nl $\mathit{generate} \: (\mathcal{P}^i)=(n,a,\hat{t}_i,E_z,E_k)$
				
				\nl $m^i=enc(\mathcal{P}^i,D^i_{\mathit{PK}})$
				
				\textbf{end \: for}} 
			\textbf{end \: for}}
		\nl return $c_l=((m^i,m^{\mathit{i+1}}),(m^j,m^{\mathit{j+1}}))$
		\textbf{end}}
	\caption{Algorithm for order creation.}
	\label{alg:crtn}
\end{algorithm}
\setlength{\textfloatsep}{0pt}

\smallskip\noindent\textbf{Step 1: Order creation: schedule creation at the owner.} The homeowner $O$ first creates a schedule, say $\mathit{Schedule}$, for device actuation. The creation of schedule is inherently specific to the preferences of the owner on a day to day basis and can include all or a subset of the home devices. The schedule creation does not require interaction with any device $D^i$ or the hub $H$. Below we show a partially ordered timeline/schedule of four devices:

\centerline{$\mathit{Schedule}=((D^1,D^2),(D^3,D^4))$}

Where only the elements of the same subset are comparable, e.g., $D^1$ with $D^2$, and, $D^3$ with $D^4$, based on the timeline. The owner converts this schedule to a verifiable device ordering (see Algorithm~\ref{alg:crtn}), before sending it to hub. The function $\mathit{create}(c_l,D^N)$ of Algorithm~\ref{alg:crtn} converts a schedule for devices into the partially ordered sets of commands of length $c_l$, where $l\subseteq N$. Line 3 considers a pair\footnote{We consider only pairs of devices, to simply demonstrate the relative ordering. However, a different subset may contain as large as the total number of devices.} of devices in the topology of the hub. Line 4 creates a mutually dependent schedule for the pair of devices with temporal dependency. Lines 5 and 6 consider all of these mutually dependent pairs, decide the state of command as $z=(s_{\mathit{on}}\vee s_{\mathit{off}})$, and then, generate a unique puzzle $\mathcal{P}^i$ for each device in Line 7. Here, the puzzle message contains a tuple of variables $(n,a,\hat{t}_i,E_z,E_k)$, where $n$ is the product of two large prime numbers $p$ and $q$, $a$ is a random number, $\hat{t}_i$ is the time-complexity of the puzzle, $E_z$ is the encrypted command $z$ using key $k$, and $E_k$ is the encrypted key $k$. Line 8 encrypts each puzzle $\mathcal{P}^i$ into a message $m^i$ for a device $D^i$ using the public key of the device $D^i_{\mathit{PK}}$. Line 9 returns an assembled order $c_l$ that contains an encrypted message for each device corresponding to the mutually dependent devices in the schedule. This ends the creation of a relative order for the chosen set of devices. Next, the homeowner sends securely this order, say $\mathit{Order}$, to the hub, as follows:

\centerline{$\mathit{Order}=(O_{\mathit{id}},H_{\mathit{id}},c_l,\mathit{Sign}(\mathcal{H},O_{\mathit{SK}}))$}

i.e., the homeowner sends its identity ($O_{\mathit{id}}$), the identity of the hub ($H_{\mathit{id}}$), and encrypted order of commands ($c_l$) along with hash digest of all three attributes.

\LinesNotNumbered
\begin{algorithm}[t]
	\DontPrintSemicolon
	\textbf{Inputs:} set of devices ($D^N$), user-defined schedule $((D^i,D^{i+1}),(D^j,D^{j+1}))$
	
	\nl{\bf Function $\mathit{chain}(D^i,D^j,\mathcal{R})$} \nllabel{ln:chng}
	\Begin{ 	
		\nl \For{$\forall (i,j,\mathcal{R}) \: \in \mathit{schedule}(D^i,D^j); \: i\mathcal{R}j=i<j$}{
			
			\nl $^nP_n(D^i)$
			
			\nl \For{$\forall \{D^i\}_{n!} \wedge i\mathcal{R}j=true$}{
				
				\nl return $^nP^{\prime}_n\{D^i\}^n_{\mathit{i=1}}$}
			
			\textbf{end \: for}}
		\textbf{end}}
	%
	\caption{Algorithm for chaining.}
	\label{alg:chng}
\end{algorithm}
\setlength{\textfloatsep}{0pt}

\smallskip\noindent\textbf{Step 2: Token generation by the hub and token delivery to devices.} On receiving the partial order of commands from the owner, the hub verifies the sender by computing a local hash digest $\mathcal{H}^\prime$ over $(O_{\mathit{id}},H_{\mathit{id}},c_l)$. Also the hub verifies the signature using $O_{\mathit{PK}}$ and compares the received hash digest $\mathcal{H}$ with the locally computed hash digest $\mathcal{H}^\prime$. If $\mathcal{H}=\mathcal{H}^\prime$ then the hub accepts this order.

After the verification of order origination, the hub creates a token that is used for order delivery (in this step) and for data collection generated by devices (step 4). The token has three fields: command field $c_l$, data field, and toggle bit string $b_{\mathit{toggle}}$. Every token field has sensitive information regarding the device activity. Therefore, we assume that the token is encrypted using a shared symmetric key $k_s$ among the devices and the controller hub.

\centerline{$\mathcal{T}=E((c_l||\mathit{Data \: field}||b_\mathit{toggle}),k_s)$}

\begin{figure}[h]
	\begin{center}
		\includegraphics[scale=0.24]{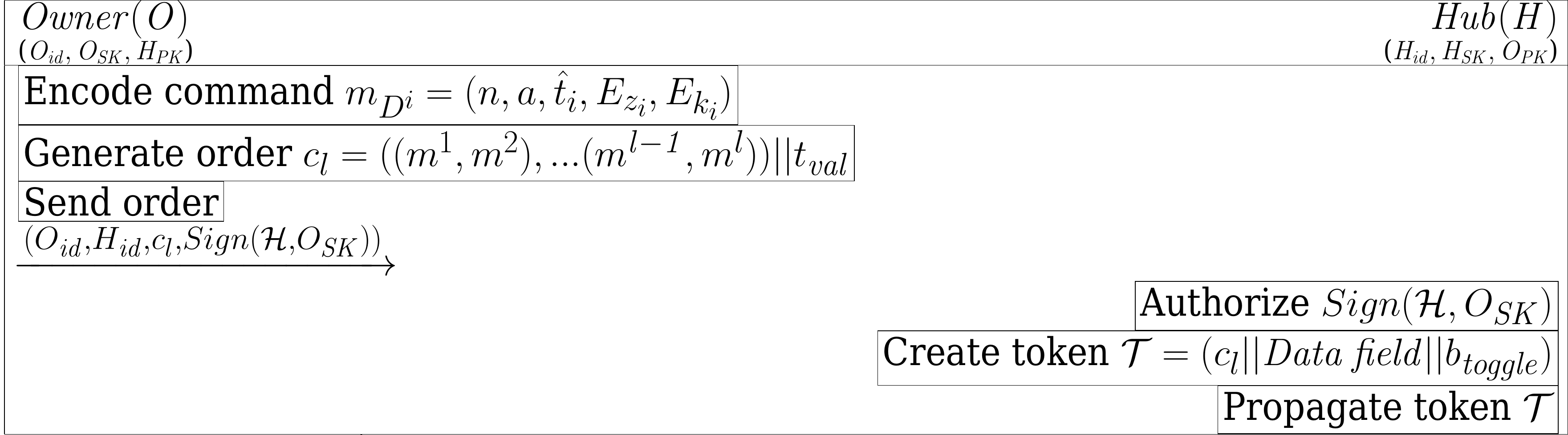}
	\end{center}

	\caption{Order creation.}
	\label{fig:s1}
\end{figure}

\begin{figure}[t]
	\begin{center}
		\includegraphics[scale=0.35]{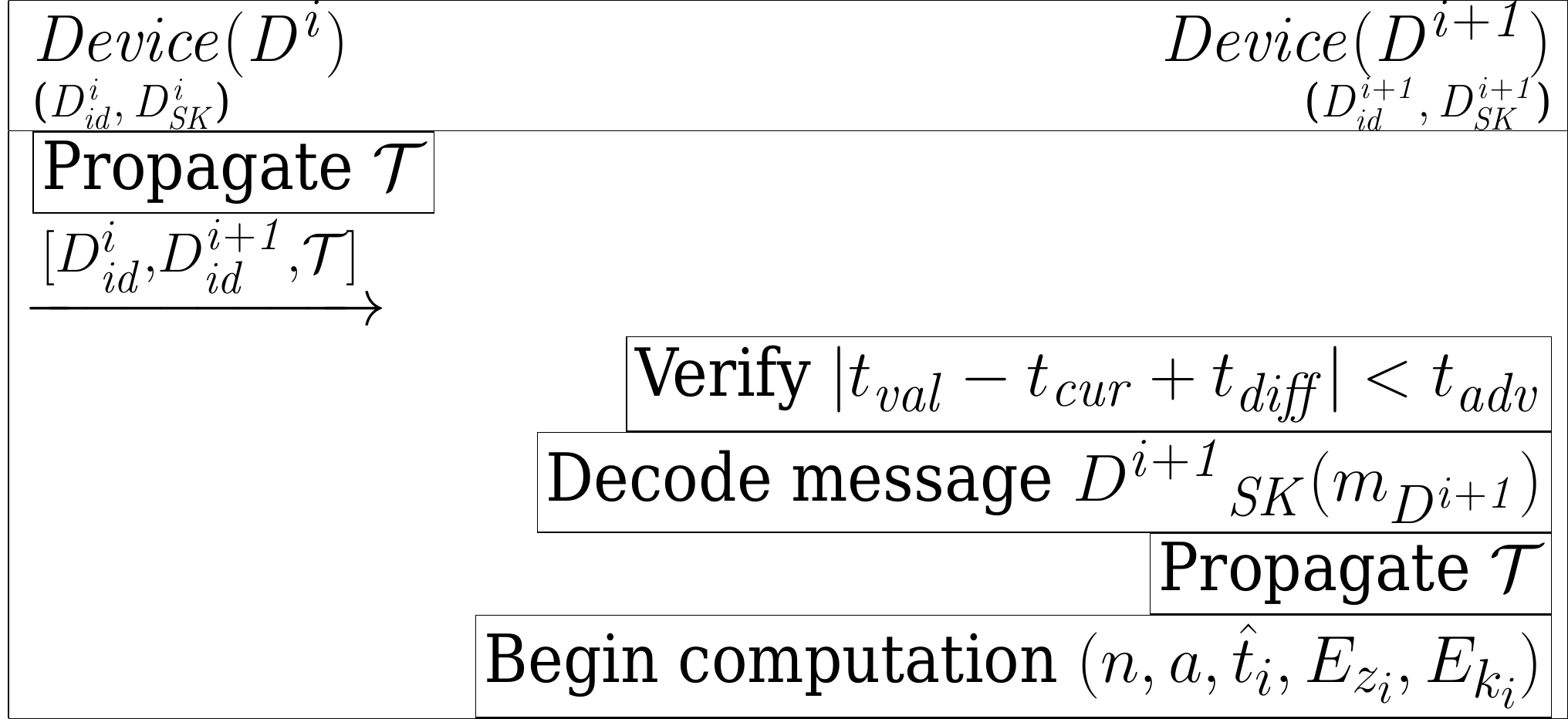}
	\end{center}
	\caption{Token circulation.}
	\label{fig:s2}
\end{figure}

The hub, then, forwards the token among all devices in the topology even if a device was not included in the schedule, as shown in Figure~\ref{fig:s1}. On receiving the token, each device decodes the corresponding command in token and forwards the token to the next peer device in the topology. The next peer device is chosen as per the topology underneath. Recall that the devices are connected in a pre-defined topology\footnote{It must be noted that from the practical deployment aspect it is difficult to connect these smart devices in a ring topology unless the devices belong to the same OEMs, e.g., Apple HomeKit. Therefore, a star or a grid topology can be used to combat the single point of failure and device heterogeneity in the current scenario.} such as in an unidirectional ring, bidirectional ring, star (fault-tolerant), grid, mesh or hybrid setting. Note that the token rotates constantly in the topology (see Figure~\ref{fig:s2}).


Algorithm~\ref{alg:chng} explains the linear ordering of devices in function $\mathit{chain}(D^i,D^j,\mathcal{R})$. The linear ordering condition requires that each device $D^i$ and $D^j$ must satisfy: the exact same mutual ordering or relation $\mathcal{R}=\{\leq\}$ as in $c_l$. The line 2 selects each pair ($i,j$) of device that is paired under relation $\mathcal{R}=\{\leq\}$ in the original schedule. Line 3 enumerates all possible permutations of these devices, say $^nP_n(D^i)$ where $i\in N$. Line 4 selects one permuted order $^nP^{\prime}_n$ of devices (from the total number of permutations) such that the precedence relation still holds true, i.e., $i\mathcal{R}j=true$. Finally, in line 5, the selected topological order $^nP^{\prime}_n$ is returned from all possible permuted orders $^nP_n(D^i)$ of the devices.

\begin{figure}
	\begin{center}
		\fbox{
			\begin{minipage}[t]{3.2in}
				\noindent
				\normalfont
				\textbf{The puzzle messages}:
				\begin{center}
					\begin{tabular}{l l l}
						$n=pq$ & & \\
						
						$\hat{t}=St^{\prime}$ & & \\
						
						$E_z=\mathit{enc}(z,k)$   & &    \\
						
						$E_k=k+{a^{2^{\hat{t}}}} \ (mod \ n)$     &    &  \\
					\end{tabular}
				\end{center}
				\noindent
				\textbf{The puzzle computation ($\mathcal{P}$)}:
				\begin{enumerate}[noitemsep,nolistsep,leftmargin=0.2in]
\item Initially, $\mathcal{P}$ receives as input a secret key $k$ and encrypt the original message $z$ denoted as $E_z$. It must be noticed that each individual command corresponding to a device $D^i$ is secured in the form of an encrypted message $m_i$. Furthermore, a cascaded command as a whole contains multiple messages of these types.
					
\item Subsequently, $\mathcal{P}$ receives inputs as a secret key $k$, random number $a$, puzzle difficulty level $\hat{t}$, modulus $n$, and then generate $E_k$. The puzzle computation relies on the secrecy of key $k$ used to encrypt a secret message $E_z=(z,k)$. Also, $\hat{t}=S*t^{\prime}$ denotes the difficulty level of the puzzle for a specific device which can perform $S$ number of square operations per second and $t^{\prime}$ is the time to decrypt the message using a regular encryption scheme.
					
\item The puzzle $\mathcal{P}$ includes the tuple ($n,a,\hat{t},E_z,E_k$) for which the computation task is to be solved. Evidently, the recipient of this puzzle would have to spend at least $\hat{t}$ amount of time to complete the computation task and reveal the key $k$.
				\end{enumerate}
			\end{minipage}
	}\end{center}
	\caption{Puzzle computation.}
	\label{fig:s5}
\end{figure}

\smallskip\noindent\textbf{Step 3: Order retrieval and puzzle computation at the devices.} Note that the activation sequence of devices is released ahead of their actual activation; however, we need to restrict them not to execute the command before prescribed time. Thus, in this step, devices perform a pre-defined computation task (as detailed in Figure~\ref{fig:s5}) to unlock and execute the owner-defined command.

It must be noted that unlocking the command is as necessary as unlocking the command within the prescribed duration, i.e., knowing the particular order of a device in the overall sequence. A device receives the computation task in the form of a puzzle as soon as the order is delivered in step 2. Subsequently, the device begins the computation task if the puzzle validity period has not expired yet. The time-bound during which a device is restricted to begin, as well as, end the computation task cannot be compressed unless the device possesses a distinguisher for the factoring problem. Therefore, the verification that the secure computation task is crucial. This can be verified through the Euler's totient as a trapdoor for factoring $n$ inside the puzzle. Note that the value of $n$ is public and the value of $\phi{(n)}$ is kept secret. Therefore, computing $\phi{(n)}$ from $n$ is as hard as integer factorization. In addition, without the knowledge of $\phi{(n)}$ the computation time for $E_k$ is directly dependent on $\hat{t}$ time-consuming square operations.


\LinesNotNumbered
\begin{algorithm}[t]
	\DontPrintSemicolon
	\textbf{Inputs:} a puzzle ($\mathcal{P}$) with a set of public variables ($n,a,\hat{t}_i,z_i,k_i$)
	
	
	\nl{\bf Function $\mathit{verify}(D^i,\hat{t}_i)$} \nllabel{ln:vrfy}
	
	\nl \Begin{
		\nl  	 \For{$\forall (i,j) \in\mathit{schedule}((D^i,D^{i+1}),(D^j,D^{j+1}))$}{
			\nl \If{${a^{2^{\hat{t}}}} \: mod \: n \equiv {a^{2^{\hat{t}} \: mod \: \phi{(n)}}} \: mod \: n$}{
				\nl \If{$(t^i_{\mathit{com}} \le t^{i+1}_{\mathit{com}}) \wedge (t^j_{\mathit{com}} \le t^{j+1}_{\mathit{com}})$}{
					
					\nl	return $\mathit{True}$}
				\textbf{end \: if}}
			\textbf{end \: for}
		}return $\mathit{False}$ \textbf{end}}
	\caption{Algorithm for delay verification.}
	\label{alg:vrfy}
\end{algorithm}
\setlength{\textfloatsep}{0pt}

Algorithm~\ref{alg:vrfy} explains the verification of time-bounded commands in function $\mathit{verify}(D^i,\hat{t}_i)$. Line 3, considers all devices that are part of the current schedule. In order to verify that a specific device has executed the command within the pre-defined interval, $O$ securely pre-computes the Euler's totient $\phi{(n)}=(\mathit{p-1})(\mathit{q-1})$ such that $p,q$ is discarded after computing the $n$ and $\phi{(n)}$. Line 4, verifies the time bound for each of these devices, such that

\centerline{${a^{2^{\hat{t}}}} \: mod \: n \equiv {a^{2^{\hat{t}} \: mod \: \phi{(n)}}} \: mod \: n$}

Line 5 compares the execution order of devices that have happened as a result of puzzle computation and returns true in line 6 if it is a total order.

\smallskip\noindent\textbf{Step 4: Data generation at the devices.} Once the devices have completed the puzzle computation, they generate the data as a result of the command execution. These home devices are bound to upload the locally generated data to the hub. In our scheme, the token contains an anonymous data field to securely transmit the device generated data to the hub. We use bitwise ($b$) XOR padding to overwrite the random data in the token data field as:

\centerline{$\mathit{Overwrite} \: \mathit{data} \: (b_o)=\mathit{Random} \: \mathit{data} \: (b_r) \xor \mathit{Generated} \: \mathit{data} \: (b_g)$}

It is cryptographically hard to distinguish the presence of random data from the device generated data as stored inside the token. Note that our threat model does not consider the ISP or DNS level threats, therefore, devices are only assumed to securely generate and anonymously dispatch the data to the hub and combat any passive learning attacks within the physical periphery of the home.

\noindent\textit{Collision:} Our token circulation strategy and the token structure are primarily for smart home scenarios, where we assume that the single token field can accommodate the peak hour traffic. However, in case the peak hour traffic exceeds and multiple devices request for data upload, for example in a multi-tenant building, then to avoid the collision situation more data fields are required. A simple approach is to create sub-fields inside the data field such that each sub-field belongs to a unique device. Thus, each device can fairly utilize the data upload capacity in any round during the token circulation.

\subsection{Time Analysis}
The time spent during the token circulation and puzzle computation is directly proportional to the number of devices connected in the network. For example, token begins at time ($t^H_{\mathit{beg}}$) at the hub and completes the first round of token circulation at time ($t^H_{\mathit{end}}$). The time spent at $i$th device is ($t^i_{\mathit{fwd}}-t^i_{\mathit{rcv}}$) that receive the token at time ($t^i_{\mathit{rcv}}$) and forward it to next device at time ($t^i_{\mathit{fwd}}$). Therefore, the total time spent in one round of token circulation:
$$t_{\mathit{sum}}=(t^H_{\mathit{end}}-t^H_{\mathit{beg}})-(t^i_{\mathit{fwd}}-t^i_{\mathit{rcv}})_{\mathit{i=1}}^N$$
Note that the token circulation time is sequenced and linear w.r.t. the number of devices. While the puzzle computation time $t^i_{com}$ varies independently among all devices. So the puzzle computation time at $i$th device is $t^i_{\mathit{com}}\approx \hat{t}^i$. In order to optimize the puzzle computation time and still retain the verifiable guarantees on the artificial delay, we consider three type of puzzles.
  \begin{itemize}[noitemsep,nolistsep,leftmargin=0.05cm]
  	\item \textit{For comparable devices:} Each pair of comparable devices in the topology requires that $\hat{t}$'s are {\em at least} $(N-1)(t^N_{\mathit{fwd}}-t^{N-1}_{\mathit{fwd}})$ apart. The devices forward the token before beginning the local computation task. Any two adjacent devices ($D^{N-1},D^N$) that begin the computation after forwarding the token, must possess:
  $$|\hat{t}^{N}-\hat{t}^{N-1}|\geq(N-1)(t^N_{\mathit{fwd}}-t^{N-1}_{\mathit{fwd}})$$
    	\item \textit{For incomparable devices:} The set of incomparable devices require that $\hat{t}$'s are {\em exactly} $(j-i)(t^j_{\mathit{fwd}}-t^{i}_{\mathit{fwd}})$ apart. Every time a device $D^i$ forwards a token to $D^{i+1}$ it jumps $(t^{i+1}_{\mathit{fwd}}-t^{i}_{\mathit{fwd}})$ ahead on the computation timeline with respect to next device due to token propagation delay. Therefore, in order to provide an identical time of actuation for all incomparable devices:
  $$|\hat{t}^{i}-\hat{t}^{j}|=(j-i)(t^j_{\mathit{fwd}}-t^{i}_{\mathit{fwd}})$$
  	\end{itemize}

The total number of slots required is $(N-k)+1$ where $k$ represents the number of comparable devices. In particular, each comparable device requires a unique and non-overlapping $|\hat{t}|$ w.r.t. other comparable devices; while each incomparable device can be scheduled for an identical and overlapping $|\hat{t}|$.

\noindent\textit{Token frequency:} The token frequency is a crucial attribute from the perspective of how early a user can decide the schedule for all $N$ devices and, how many data upload requests are received during the peak hours. The frequency of token circulation can be either {\em fixed} or {\em random}. Let us assume a {\em fixed} slot $i$ between any two consecutive rounds of the token circulation such that the token begins a new round at every $i$th unit of time. The optimal length of the slot is the same as the maximum $\hat{t}^i$ in any schedule.
$$\mathit{Slot \: length}=max\{\hat{t}^i\}_{i=1}^{l\in N}$$
For example, if $(\hat{t}^1,\hat{t}^2,\dots,\hat{t}^{l-1},\hat{t}^l)$ is the time-bound for $l$ devices in any scheduled workflow then the slot length is same as the farthest possible device on the timeline of a scheduled workflow. 


   \begin{table*}[t]
   	\BB
 	\begin{center}
 	\begin{tabular}{l l l l}
 	\hline
 	{Protocol} & {Cost at device} & {Cost at hub} & {Cost at owner}  \\ \hline
 	Proposed scheme & ($2D+1XOR+\hat{t}Me$) & ($1S+1E+1XOR$) & ($1S+lE+\hat{t}mod{\phi{(n)}}Me$) \\
 	Scheme~\cite{tifs} & ($3H+7XOR+1E+1D$) & ($5H+8XOR+1E+1D$) & - \\
 	Scheme~\cite{shen} & ($3M+2H+4XOR$) & ($1H+4XOR$) & - \\
 	Scheme~\cite{thir} & - & ($Cp_m+Tp_m^{\prime}$) & - \\ \hline
 	\end{tabular}
 	\caption{Cost comparison between our scheme and existing schemes~\cite{tifs,shen,thir}.}
 	\label{table:cost}
 	\BBB\BBB
 	\end{center}
 \end{table*}

Table~\ref{table:cost} represents the cost comparison based on mathematical operations such as encryption ($E$), decryption ($D$), signature generation and verification ($S$), exclusive-OR ($XOR$), hashing ($H$), scalar multiplication ($M$), and modular exponentiation ($Me$). The scheme in~\cite{thir} imposes a relative overhead such that ($p_m$) number of masking packets are required per traffic flow in case the traffic flow is lesser than a pre-defined threshold value. Therefore, the total overhead per traffic flow is ($Cp_m$) where $C$ is the communication overhead per packet. Similarly, if the traffic flow is above the threshold value then those excess packets $p_m^{\prime}$ are delayed and stored inside a queue. Therefore, the total latency per traffic flow is ($Tp_m^{\prime}$) where $T$ is the latency overhead per packet. As shown here, that our proposed scheme requires the minimum number of operations. Further, in our approach the computational complexity at devices is variable and it depends on the required number of modular exponentiations, e.g., $\hat{t}$, as initialized by the owner.

\section{Security Analysis}\label{sec:analysis}
This section provides the security analysis for proposed scheme. We first model the security experiment, below, like the standard security model.

  \begin{attackgame} Let $\mathcal{I}$ be the order-preserving protocol between the challenger and adversary $\mathcal{A}$ then the attack game works as:
  \end{attackgame}
  \begin{itemize}[noitemsep,nolistsep,leftmargin=0.05cm]
  	\item Public parameter generation: The challenger generates $(n,a)$ using $\mathit{param}(1^{\lambda})$.
  	\item Puzzle generation: The challenger generates $\mathcal{P}$ using $\mathit{puzgen}(n, a,$ $\hat{t}, E_k, E_z)$.
  	\item Query phase: An adversary attempts to attack $\mathcal{I}$ through token query given the access to a recently generated token $\mathcal{T}^{\prime}$. The adversary sends a value $\hat{t}$ to the challenger. The challenger generates the corresponding puzzle $\mathcal{P}$ and adds in $c_l$. The follower devices receive $c_l$, extract the unique puzzle, and execute the command. The challenger then sends $\mathcal{T}^{\prime}$ to the adversary.
  	\item State identification attempt: The adversary attempts to retrieve the intermediate state of the computation task and attempts to solve the puzzle earlier than device through $t^{\mathcal{A}}_{\mathit{com}}$ for the same $\mathcal{P}$ in $\mathcal{T}^{\prime}$, such that $t^{\mathcal{A}}_{\mathit{com}}$ is lower than the original $\hat{t}$.
  \end{itemize}
An adversary $\mathcal{A}$ wins the game, if $t^{\mathcal{A}}_{\mathit{com}}<\hat{t}_i$ and the owner outputs $\mathit{accept}$. The probabilistic advantage of the adversary, $\mathit{Adv(\mathcal{A})}$, for winning the game is:
$$\mathit{Adv(\mathcal{A})}=Pr[t^{\mathcal{A}}_{\mathit{com}}<\hat{t}_i]$$

We present a sequence of games as $\texttt{Game 0}$ to $\texttt{Game 2}$. Each $\texttt{Game i}$ shows that the advantage of an adversary $Pr[t^{\mathcal{A}}_{\mathit{com}}<\hat{t}_i]$ is negligibly small. Similarly, each subsequent game $\texttt{Game (i+1)}$ is produced through previous game such that the changes in secret parameters remain indistinguishable to the adversary. Therefore, the advantage of an adversary through changes in secret parameters (i.e., transitioning from one game to another) remains negligibly small. If $Pr[\mathcal{A}(i) \rightarrow 0]-Pr[\mathcal{A}(i+1) \rightarrow 0]$ is non-negligible then that adversary can be used as a distinguisher or as a solver for the integer factorization problem in our scheme; where $Pr[\mathcal{A}(i)]$ and $Pr[\mathcal{A}(i+1)]$ represent the probability of adversary winning the $\texttt{Game i}$ and $\texttt{Game (i+1)}$, respectively. The $\texttt{Game 0}$ represents the original attack such that $(t^{\mathcal{A}}_{\mathit{com}}=\hat{t}_i)$ and the artificial delay before the command execution is kept null. The $\texttt{Game 1}$ represents the attack with enhanced $t^{\prime}$ while $(t^{\mathcal{A}}_{\mathit{com}}<t^\prime\wedge\hat{t}_i)$. Similarly, the $\texttt{Game 2}$ represents the attack with general $t^{\prime}$ while $(t^{\mathcal{A}}_{\mathit{com}}=t^\prime)$ but $(t^{\mathcal{A}}_{\mathit{com}}<\hat{t}_i)$.


  \noindent\texttt{Game 0:} [\textbf{Record attack}] Let us assume that the puzzle $P_1$ contains $\hat{t}=0$ then the device must perform only one iteration to compute and decode the enciphered command. However, an adversary cannot distinguish an early puzzle such as $P_1$ from a delayed puzzle such as $P_2$ for which $\hat{t}>0$. In the token query phase, an adversary gathers the token transcripts for a known value of $\hat{t}$.

  {\upshape
  	\small
  	\ttfamily
  	\begin{tabbing}
  	Exp\=eri\=ment $\aexp^{t^{\mathcal{A}}_{\mathit{com}}=\hat{t}_i}$\\
  	\> let $c_l((m^1,m^2),\ldots,(m^{l-1},m^l)) \leftarrow \aaa(\mathcal{T}) $\\
  	\> generate $m^i(\hat{t})$ at random\\
  	\> $(m^k(\hat{t})) \leftarrow \aaa(c_l((m^1,m^2),\ldots,(m^{l-1},m^l)))$\\
  	\>if ($m^k(\hat{t})=m^i(\hat{t})$) \\
  	\> \> return $1$\\
  	\> else return $0$
  	\end{tabbing}
  }

It is computationally hard to distinguish between encrypted commands and to identify the command that carries known $\hat{t}$ within time $t^{\mathcal{A}}_{\mathit{com}}=\hat{t}_i$. An adversary can distinguish the commands with the advantage
$$\mathit{Adv(\mathcal{A})}_{\texttt{Game 0}}=Pr[m^i(\hat{t})\leftarrow c_l((m^1,m^2),\ldots,(m^{l-1},m^l))]\leq \epsilon$$

  \noindent\texttt{Game 1:} [\textbf{Clone attack with lesser $t^{\prime}$}] Let us assume that the adversary computes the puzzle in time $t^{\mathcal{A}}_{\mathit{com}}<t^\prime \wedge \hat{t}_i$ where the unit time capacity $t^{\prime}$ of the adversary is slower. Therefore, the advantage of the adversary depends on the probability to compute the $\hat{t}$ square operations faster than the home device. This requires that the adversary can solve the prime factors for modulus $n$.
  $$\mathit{Adv(\mathcal{A})}_{\texttt{Game 1}}=Pr[(p,q)\leftarrow(n)]\leq \epsilon$$   

  \noindent\texttt{Game 2:} [\textbf{Clone attack with general $t^{\prime}$}] Let us assume that the adversary can compute as fast as the home device, i.e., $t^{\prime}$. In particular, the adversary can also perform $S$ number of square operations per second. The probability $Pr[t^\prime=t^{\mathcal{A}}_{\mathit{com}}<\hat{t}_i]$ that an adversary $\mathcal{A}$ can solve a puzzle with the difficulty level $\hat{t}$ in lesser time than the home device is negligibly small. Since the adversary must compute $S^{\prime}$ number of operations for each $S$ operations at home device where ($S^{\prime}-S>\epsilon$):
     $$\mathit{Adv(\mathcal{A})}_{\texttt{Game 2}}=Pr[t^{\mathcal{A}}_{\mathit{com}}(S^{\prime}t^{\prime})\leftarrow \hat{t}_i(St^{\prime})]\leq \epsilon$$     

In this sequence of games $\texttt{Game 0}$ through $\texttt{Game 2}$ the total advantage of the adversary $\mathit{Adv(\mathcal{A})}$ depends on the sum of the probability to win each of these games.
$$Pr[t^{\mathcal{A}}_{\mathit{com}}=\hat{t}_i]+Pr[t^{\mathcal{A}}_{\mathit{com}}<t^\prime \wedge \hat{t}_i]+Pr[t^\prime=t^{\mathcal{A}}_{\mathit{com}}<\hat{t}_i]\leq \epsilon$$
Overall, the advantage of the adversary is proportional to the availability of computational resources to solve the puzzle for all devices in parallel. Similarly, the advantage of the adversary with respect to a single puzzle and a single device are proportional to the availability of computational resources to solve the inherently sequential operations of that individual puzzle. Therefore, the total advantage of an adversary to clone the entire timeline depends on the total computational power for both, the parallel and the sequential operations to decrypt the command ahead of time.

\begin{table}[t]
	\begin{center}
	\begin{tabular}{l l l l l l}
	\hline
	{Properties} & \cite{tifs} & \cite{shen} & \cite{thir} & Our scheme  \\ \hline
	Upstream direction & $\checkmark$ & $\checkmark$ & $\checkmark$ & $\checkmark$ \\
	Downstream direction & $\checkmark$ & $\times$ & $\checkmark$ & $\checkmark$ \\
	Verifiable delay & $\times$ & $\times$ & $\times$ & $\checkmark$ \\
	Partial ordering & $\times$ & $\times$ & $\times$ & $\checkmark$ \\
	Total ordering & $\times$ & $\times$ & $\times$ & $\checkmark$ \\
	Privacy & $\times$ & $\times$ & $\checkmark$ & $\checkmark$  \\
	Passive attack resistant & $\times$ & $\times$ & $\checkmark$ & $\checkmark$ \\
	Active attack resistant & $\checkmark$ & $\checkmark$ & $\times$ & $\checkmark$ \\ \hline
	\end{tabular}
	\caption{Comparison between our and existing schemes.}
	\label{table:cmp}
	\end{center}
\end{table}

In Table~\ref{table:cmp} a comparison is shown between our proposed scheme and the existing work. The comparison is based on the data flow direction, ordering, verification of ordering, privacy violation, and attack resistance.


\section{Experimental Evaluation}
\label{sec:Experimental Evaluation}
This section evaluates our proposed system using our prototype implementation. We describe the mock-up testing IoT application, experimental setup and overall results from our experiments.

\subsection{Experiment Setup}
\begin{figure}[t]
\BBB
	\centering
	\includegraphics[scale=0.2]{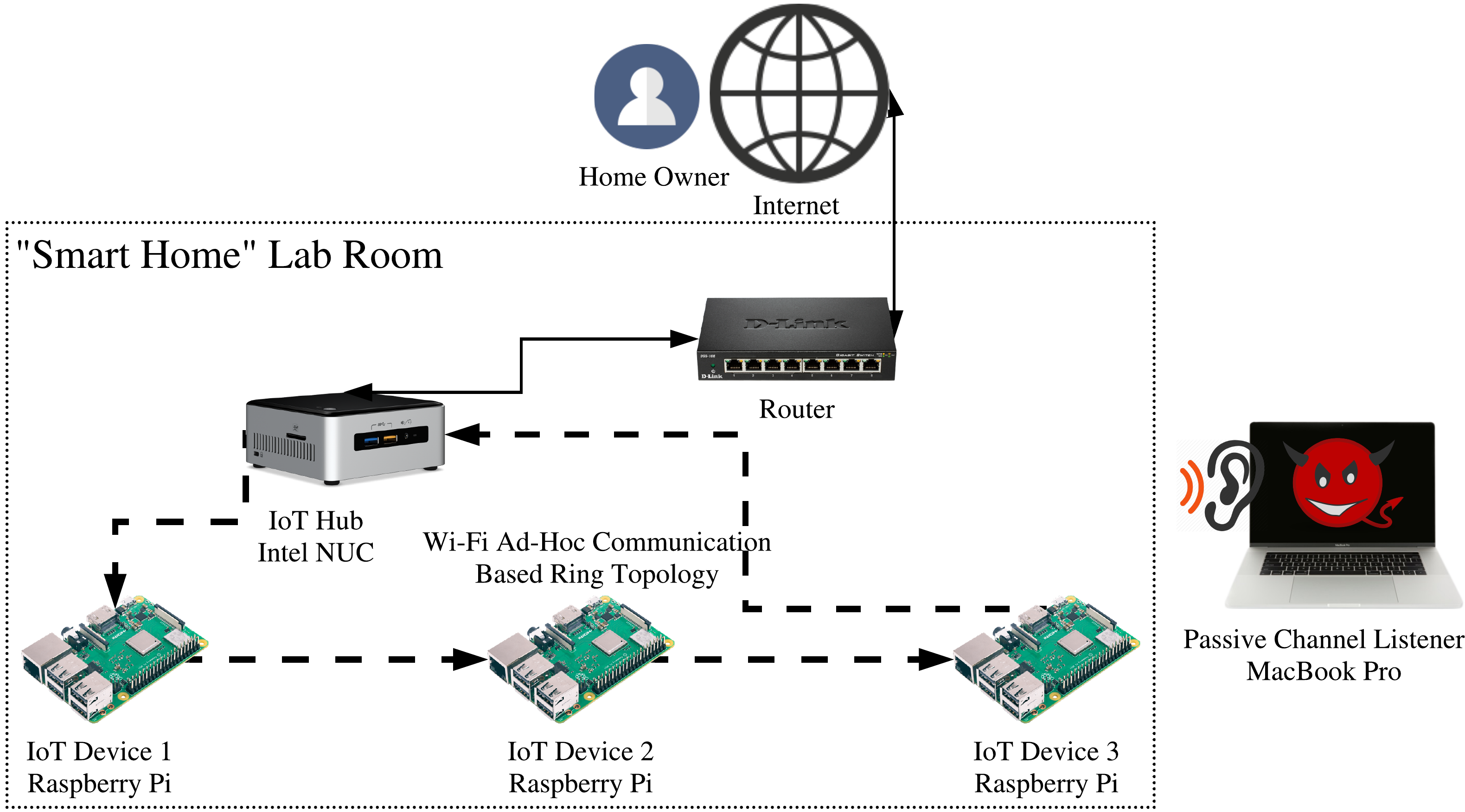}
	\caption{Experimental testbed in our lab}
	\label{fig:testbed}
\end{figure}

\B
In order to demonstrate the effectiveness and performance of our proposed architecture, we developed the prototype implementation and setup the testbed in our lab (as shown in Figure~\ref{fig:testbed}). This proof-of-concept prototype implements the protocols described in Section~\ref{sec:decoup} and a test application with Python. In this mock-up IoT application, a device awaits and executes two types of command given by the homeowner. The ``Set'' command will change a variable in the program of the target device while the ``Read'' command will require the device to send the variable together with device status, e.g., RAM and CPU usages, back to the homeowner. This mock-up application is created to simulate two-way communication between the homeowner and devices. Further, note that the application runs on top of the approach, we proposed in this paper.

Figure \ref{fig:testbed} depicts the architecture and configuration of the smart home testbed. An Intel NUC system is programmed to work as the hub $H$ that forwards information between the homeowner and smart devices. The hub $H$ equips with two network interfaces: (\textit{i}) Ethernet interface that has connections to receive the homeowner-defined schedule and sends the data of smart devices to the homeowner, and (\textit{ii}) Wi-Fi interface that is used to communicate with smart home devices. Smart home device programs are deployed on three Raspberry Pis (${D^1, D^2, D^3}$) (3rd Gen B+ Model) that are equipped with built-in Wi-Fi interface. Wi-Fi interfaces on the hub $H$ and the devices ($D^1, D^2, D^3$) are configured to work in Wi-Fi ad-hoc mode~\cite{Anastasihh}, which enables direct device-to-device communication. All Wi-Fi interfaces are configured with pre-defined Wi-Fi channel, static IP address, and routing information to have a ring topology. A MacBook Pro with its Wi-Fi interface is deployed in a different room next to the testbed performs as an adversary, who listens and dumps all channel activities on the pre-defined channel into the pcap (packet capture) file for future analysis.

\BBB
\subsection{Results}
\BB
Based on the testbed described above, we performed different experiments to evaluate the proposed system and the approach. We first validate our system to check whether it could prevent the adversary from learning device activity from channel activity or not. Then, we explore the performance of the proposed ring topology communication with a set of experiments.

\noindent\textbf{Decoupling channel activities from device activities.} To evaluate the effectiveness that our system protects against the passive channel listeners, we first defined a sequence of sample user commands, e.g., $D^1$: Set, $D^2$: Set, $D^2$: Read, and $D^3$: Read. In a one-minute experiment, these commands will be issued in 10 seconds intervals. We performed the experiment by executing the above-mentioned sequence in our proposed system and also over Wi-Fi infrastructure network without a ring topology to compare with. The channel activities are recorded by the passive channel listener laptop deployed near our testbed.

\begin{figure}[t]
\BBB
\begin{center}
  \begin{minipage}[b]{.99\linewidth}
  \centering
  \includegraphics[scale=0.5]{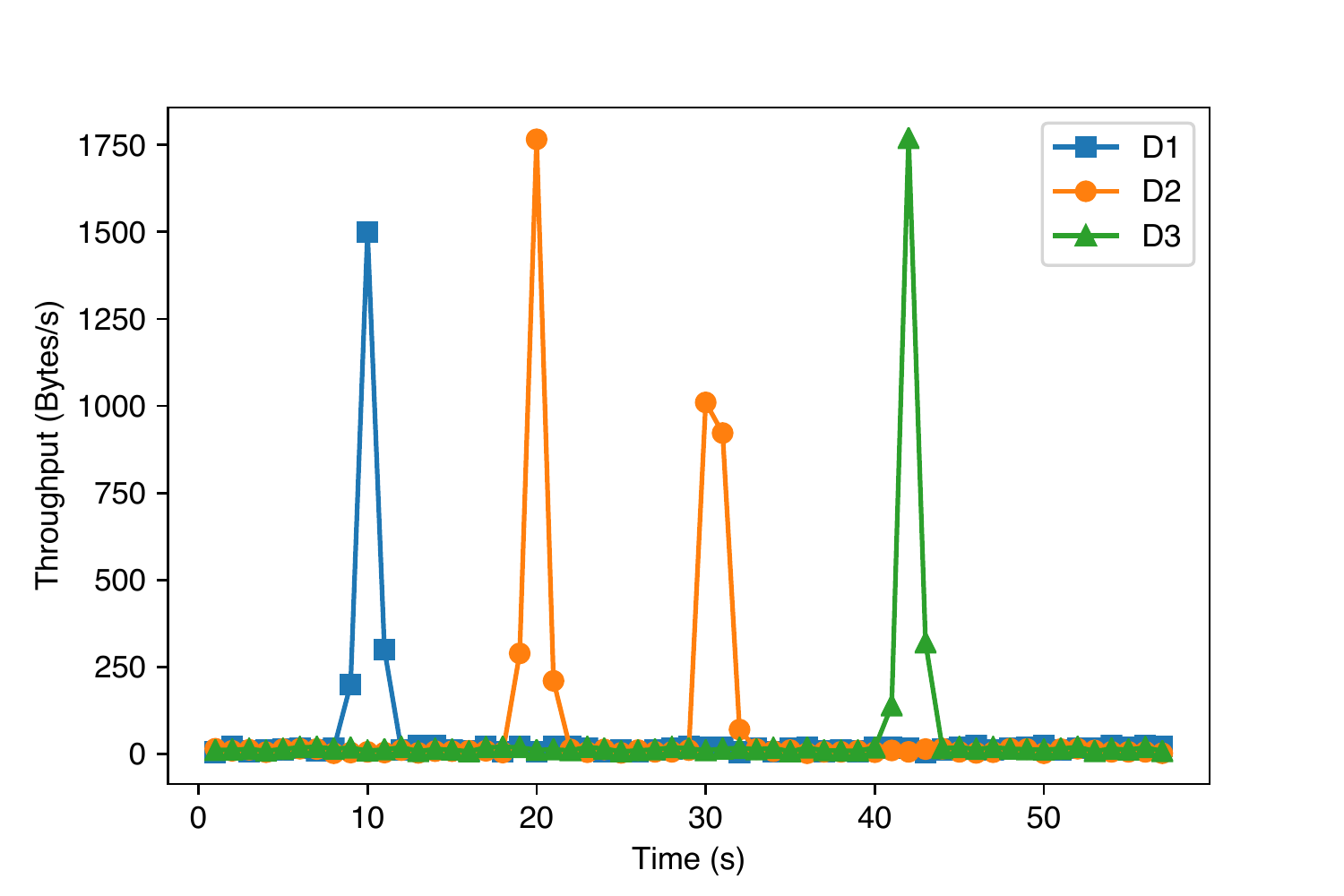}
  \subcaption{Devices working in common IoT settings}
  \label{fig:infra}
  \end{minipage}
  \begin{minipage}[b]{.99\linewidth}
  \centering
    \includegraphics[scale=0.5]{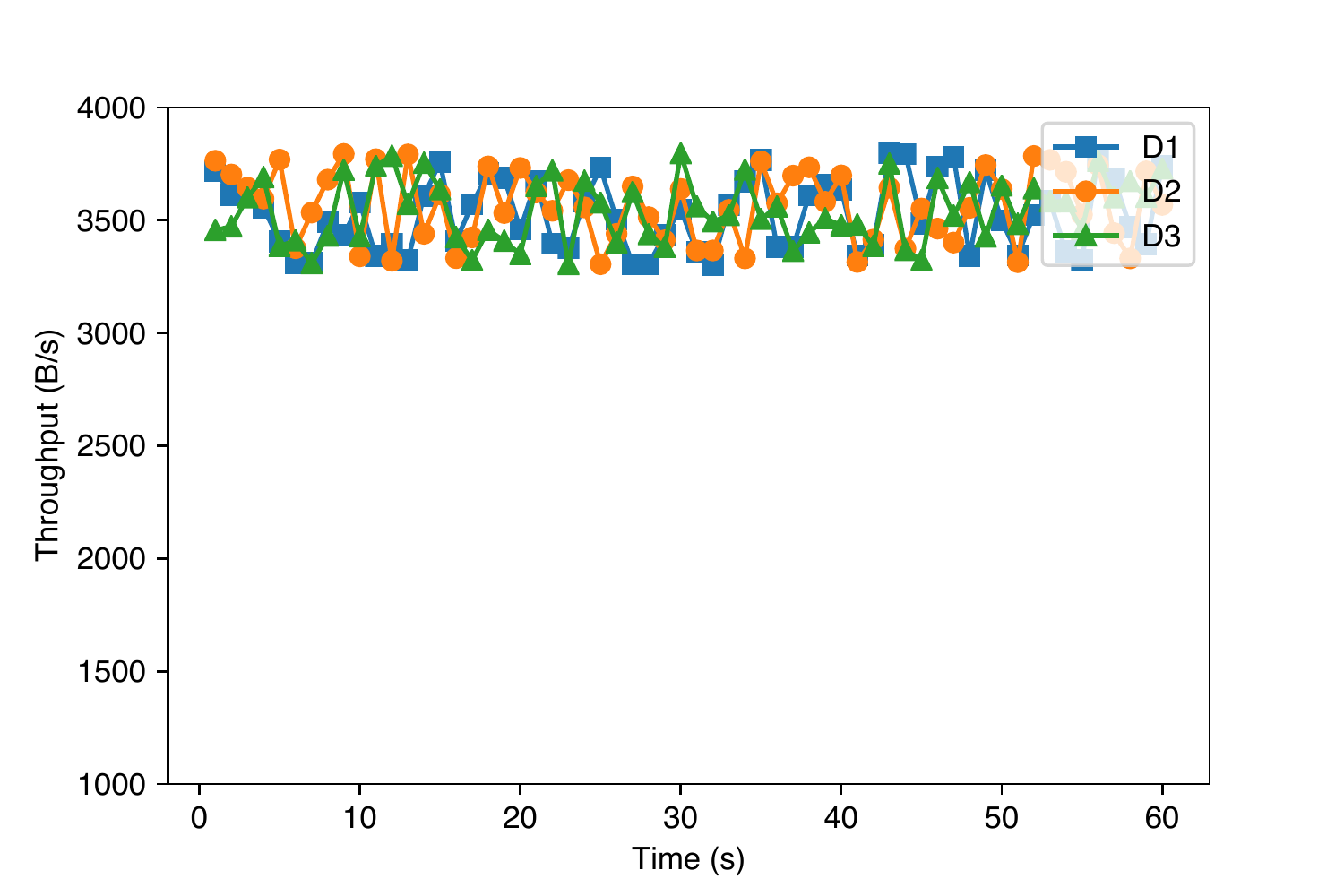}
    \subcaption{Devices working with our proposed system}
    \label{fig:adhoc}
  \end{minipage}
 \end{center}
\BB
\caption{The adversarial view due to observing channel activities}
\label{fig:channel}
\end{figure}

Figure \ref{fig:infra} shows the passive adversary's view, i.e., which device receives a message from the homeowner at which time due to channel activities in the experiment. It is clear that each time the user sends a command to devices or devices send data to the user, there will be a peak in the channel activity. Thus, the adversary infers the device activity and user-device interaction from channel activity. In contrast, Figure~\ref{fig:adhoc} shows the effectiveness of our proposed approach. Note that the channel activity patterns are completely eliminated, due to the token ring communication. 

\noindent\textbf{Communication latency.} Instead of sending individual commands or data to/from devices or hub, the commands and data in our system are encapsulated in tokens and transmitted in a ring topology, which will incur additional communication latency. We performed experiments to evaluate the impact of increasing latency as the number of devices in the ring topology increases. Since we only have a very limited number of \emph{real} devices, we modified our protocol to simulate the scenario that includes a large number of devices to investigate an impact on communication latency. To achieve this goal, we add a counter in each token. When the hub generates the token, it sets the counter equals to the number of devices we want to simulate in the experiments. This counter is decreased by one when each device receives and forwards the token to the next device. When the last device ($D^3$ in our testbed) in the ring topology receives the token, it checks the value of the current counter. If the value of the counter is more than zero, $D^3$ forwards the token to the first device ($D^1$) to extend the ring topology. If the counter number is less than or equal to zero, the last device forwards the token back to the hub to complete a single round of the token. Here, since each device will receive the same token multiple times, we also need to prevent the device from solving same puzzles and executing same commands multiple times. To do so, a unique token ID is added to each token. Thus, the device solves the puzzle and executes the command only when the device gets a new token ID.

\begin{figure}[t]
\BB
    \centering
    \includegraphics[scale=0.4]{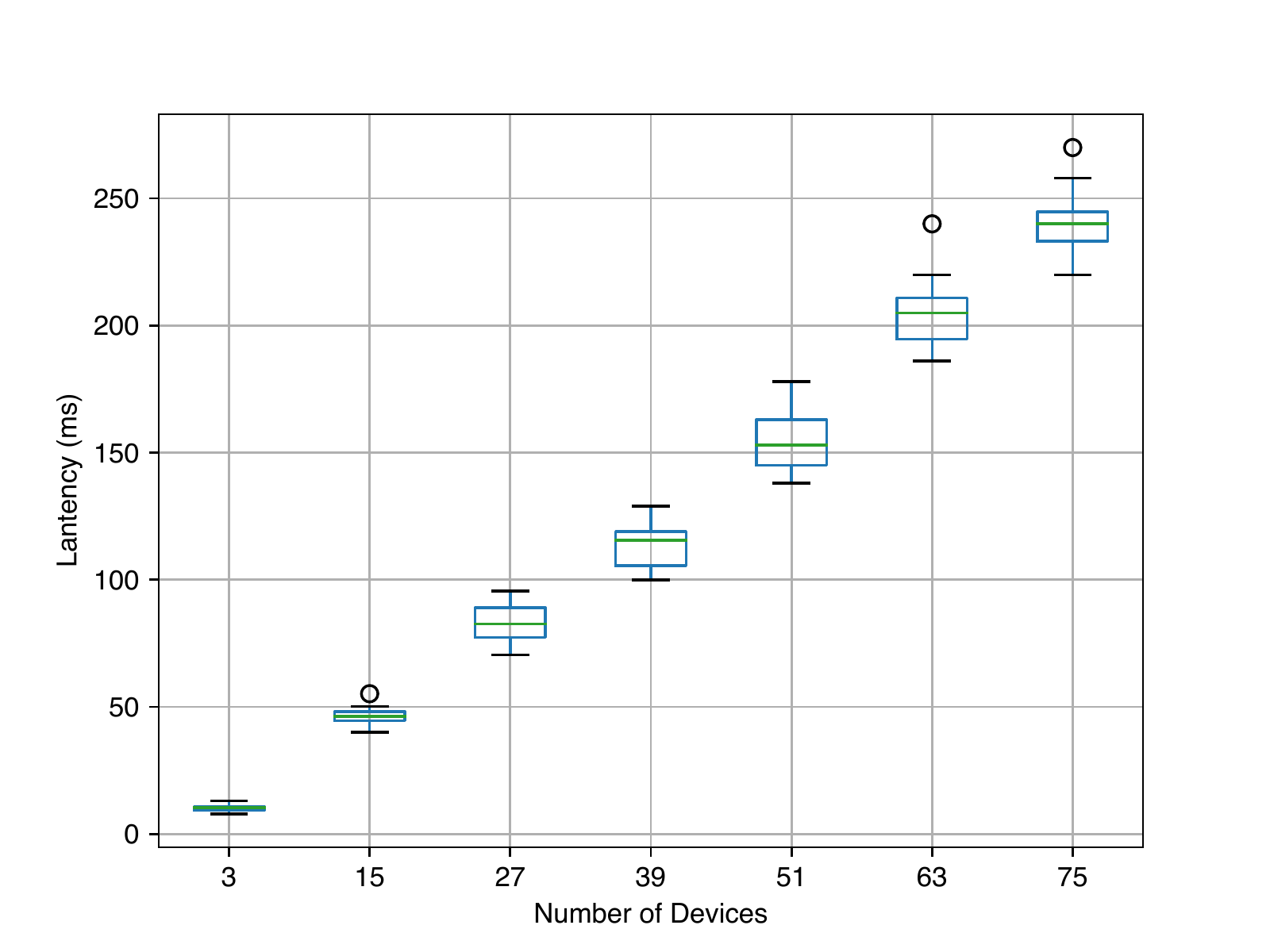}
    \BB
    \caption{Impact of ring topology on the latency}
    \label{fig:latency_exp}
\end{figure}

Figure~\ref{fig:latency_exp} shows that both mean and variation of the latency increase as more devices added into the ring topology. The mean of latency rises linearly at the beginning as each additional hop in the ring topology introduces more latency. After around 39 devices in the ring topology, the mean latency starts rising faster, since the length of token also increases with the growth of a number of devices. Consequently, it may take more time at devices to transmit the token to the next hop. Of course, the ring latency is not affected, when there is a few devices, since the number of commands and toggle bit strings decrease in the token as an decreasing number of devices. The variation of latency gets larger since it is more likely that the token transmission at more hops get delayed or re-transmitted because of unexpected interference or system lag at the devices.

\begin{table}[h]
\BBB
\begin{tabular}{|l|l|l|l|l|l|}
\hline
\textbf{\scriptsize \# Devices}       & {\scriptsize 3}       & {\scriptsize 27}      & {\scriptsize 51}   & {\scriptsize 63}   & {\scriptsize 75}    \\ \hline
\textbf{\scriptsize Avg. Token Len(bytes)} & {\scriptsize 1807}  & {\scriptsize 5024}  & {\scriptsize 8189} & {\scriptsize 9757} & {\scriptsize 11305} \\ \hline
\end{tabular}
\caption{Average token length for ring topology size}
\label{tbl:toklen}
\BBB\BBB\BBB
\end{table}

\noindent\textbf{Communication and computation overheads.} It is also important to understand the additional overheads incurred due to our approach. We first measure how the length of token increases as the number of devices grows. Table~\ref{tbl:toklen} shows that the average token length grows linearly. We also use a USB power meter to measure the power consumption of the Raspberry Pi in different working states. Table \ref{tbl:energy} shows that our system introduces 63\% more power consumption to completely eliminate the channel activity patterns of all devices.

\begin{table}[t]
\BB
\begin{tabular}{lllll}
\cline{1-4}
\multicolumn{1}{|r|}{\textbf{\scriptsize State}}                   & \multicolumn{1}{c|}{\scriptsize Idle}  & \multicolumn{1}{c|}{\scriptsize IoT App w/o Ring Sys.} & \multicolumn{1}{c|}{\scriptsize IoT App w/ Ring Sys.} &  \\ \cline{1-4}
\multicolumn{1}{|r|}{\textbf{\scriptsize Avg. Power}} & \multicolumn{1}{c|}{\scriptsize 2.25W} & \multicolumn{1}{c|}{\scriptsize 3.03W}                 & \multicolumn{1}{c|}{\scriptsize 4.91W}                 &  \\ \cline{1-4}
                                                       &                           &                                           &                                           &  \\
                                                       &                           &                                           &                                           &

\end{tabular}
\BBB\BBB\B
\caption{Average energy consumption of devices in different working states}
\label{tbl:energy}
\BBB\BB
\end{table}

\section{Related Work}\label{sec:rel}
There exist various IoT frameworks such as {\em Apple HomeKit}, {\em SmartThing}, {\em Azure IoT Suite}, {\em IBM Watson} IoT platform, {\em Brillo/Weave} platform by Google, {\em Calvin} IoT platform by Ericsson, {\em ARM mbed} IoT platform, {\em Kura} IoT project by Eclipse, interested readers may refer~\cite{doan,fram} for more details. Overall, the smart home communication~\cite{tam,goto} is a network component that executes commands based on contextual factors. To the best of our knowledge, none of the work~\cite{firs,thir} highlight the significance of secure device ordering in a smart home scenario. We highlight the presence of a pattern among smart home devices such that a partial ordering on device activity is observed on daily basis. The authors in~\cite{thir} have presented a privacy-preserving traffic shaping scheme to mask the channel activity and thereby the device or user activity at the ISP level. According to the scheme, if the shaped traffic rate is lower than the device traffic then the packets are queued, and if the shaped traffic rate is higher than the device traffic then the dummy packets are added to cover the original traffic rate variations. However, these techniques do not avoid the inferences on device activity pattern due to the straightforward binding between the channel activity and device activity. Our scheme decouples the channel activity from the device activity such that a communication activity over the channel at any given time cannot be coupled with a specific device activity or the user activity. The recent work~\cite{tifs,shen} provides a security framework for home devices and guarantees message anonymity and unlinkability during the communication sessions from hub to the device. The scheme is based on authentication and one-time session key agreement in a 3-way handshake protocol. However, as mentioned, authentication and encrypted message cannot prevent the inferences over the communication activity.

\section{Conclusion} This paper focuses on the security and privacy challenges in smart homes that facilitate the execution of multiple workflows. In particular, our solution avoids any inference attacks regarding these workflows that can reveal the user activity pattern, both in the past and the future. The primary source of these inference attacks is the ability to learn communication patterns among the smart home devices. The channel activity and the corresponding device activities are also sensitive from the user's privacy perspective. Therefore, decoupling the channel activity from the device activity is essential to hide the execution of scheduled workflows.

\end{document}